\documentclass[twocolumn]{aa}
\usepackage{graphicx}
\usepackage{natbib}

\usepackage{txfonts}
\begin{document}
   \title{On the origin of [NeII]~12.81~$\mu$m emission from pre-main sequence stars: Disks, jets, and accretion}


   \author{Manuel G\"udel
          \inst{1,2,3,4}
          \and
          Fred Lahuis
          \inst{5,4}
          \and
          Kevin R. Briggs
          \inst{2}
          \and
          John Carr
          \inst{6}
          \and
          Alfred  E. Glassgold
          \inst{7}
          \and
          Thomas Henning
          \inst{3}
          \and
          Joan R. Najita
          \inst{8}
          \and
          Roy van Boekel
          \inst{3}
          \and
          Ewine F. van Dishoeck
          \inst{4,9}
           }

   \offprints{M. G\"udel}

    \authorrunning{M. G\"udel et al.}
    \titlerunning{[Ne\,{\sc ii}] emission from pre-main sequence stars}
    \institute{University of Vienna, 
               Department of Astronomy,
	       T\"urkenschanzstrasse 17,
	       1180 Vienna, Austria\\
	       \email{manuel.guedel@univie.ac.at}
	  \and
	      ETH Zurich, Institute of Astronomy,
              8093 Zurich, Switzerland
         \and
             Max-Planck-Institute for Astronomy, K\"onigstuhl 17,  
             69117 Heidelberg, Germany
         \and
             Leiden Observatory, Leiden University, PO Box 9513, 2300 RA Leiden, 
	     The Netherlands 
          \and
             SRON Netherlands Institute for Space Research, P.O. Box 800, 9700 AV Groningen, The Netherlands
	  \and
	     Naval Research Laboratory, Code 7213, Washington, DC 20375, USA
	  \and
	     University of California at Berkeley, Berkeley, CA, 94720, USA
	  \and   
	     National Optical Astronomy Observatory, 950 N. Cherry Ave., Tucson, AZ 85719, USA     
	  \and   
	     Max-Planck Institut f\"ur Extraterrestrische Physik (MPE), Giessenbachstr. 1, 85748 Garching, Germany     
          }

   \date{Received 2009; accepted 2010}

   \abstract
   {Extreme-ultraviolet (EUV) and X-ray photons are powerful ionization and heating agents that drive disk chemistry, disk instabilities,
   and photoevaporative flows. The mid-infrared  fine-structure line
   of [Ne\,{\sc ii}] at 12.81~$\mu$m has been proposed to trace gas in disk surface layers heated and ionized by
   stellar X-ray and EUV radiation. } 
   {We aim at locating the origin of [Ne\,{\sc ii}] line emission in circumstellar environments by studying distributions of 
   [Ne\,{\sc ii}] emission and correlating the inferred [Ne\,{\sc ii}] luminosities, $L_{\rm [Ne\,II]}$, with stellar and circumstellar 
   disk parameters. 
   }
   {We have conducted a study of [Ne\,{\sc ii}] line emission  based on a sample of 92 pre-main sequence stars mostly belonging to
   the infrared Class II, but including 13 accreting transition disk objects, and also 14 objects that drive known jets and outflows.}
   {We find several significant correlations between $L_{\rm [Ne\,II]}$ and stellar parameters, in particular $L_{\rm X}$ and the wind 
   mass loss rate, $\dot{M}_{\rm loss}$. Most correlations are, however, strongly dominated by systematic scatter of unknown origin. 
   While there is a positive correlation between $L_{\rm [Ne\,II]}$ and $L_{\rm X}$, the stellar mass accretion rate, $\dot{M}_{\rm acc}$,
   induces a correlation
   only if we combine the largely different subsets of jet sources and stars without jets. Our results indeed suggest that $L_{\rm [Ne\,II]}$
   is bi-modally distributed, with separate distributions for the two subsamples.
   The jet sources show systematically higher
   $L_{\rm [Ne\,II]}$, by 1-2 orders of magnitude with respect to objects without jets. Jet-driving stars also tend to show higher mass
   accretion rates. We therefore hypothesize that the trend with $\dot{M}_{\rm acc}$ only reflects a trend with
   $\dot{M}_{\rm loss}$ that is more physically relevant for [Ne\,{\sc ii}] emission.
   }
   {
   The [Ne\,{\sc ii}] luminosities measured for objects without known outflows and jets are found to agree with 
   simplified calculations of [Ne\,{\sc ii}] emission from disk surface layers if the measured stellar X-rays 
   are responsible for heating and ionizing of the gas. The large scatter in $L_{\rm [Ne\,II]}$ may be introduced by variations 
   of disk properties and the irradiation spectrum, as previously suggested. If these additional factors can be sufficiently
   well constrained,  then the [Ne\,{\sc ii}] 12.81$\mu$m line should be an important diagnostic for disk surface ionization and
   heating, at least in the inner disk region. This applies in particular to transition disks also included in our sample.
   The systematically enhanced [Ne\,{\sc ii}] flux from jet sources clearly suggests a role for the jets themselves, as 
   previously demonstrated by a spatially resolved observation of the  outflow  system in the T~Tau triple. 
   }
   
   \keywords{Stars: formation --
             Stars: pre-main sequence --
	     Protoplanetary disks
             }

   \maketitle
%

\section{Introduction}
Circumstellar dust disks are the most important element in the process of planet formation. 
They have been well studied in resolved images, through the mid-infrared 10$\mu$m silicate feature, 
and through continuum emission all the way from the near-infrared to the millimeter domain.
Such studies have demonstrated ongoing grain growth and settling of dust to the disk mid-plane 
(e.g., \citealt{boekel03}) as well as the presence of inner dust-disk holes (e.g., \citealt{bouwman03}).

However, dust contributes only about 1\% to the total disk mass. The gas in which the dust is immersed 
is much more challenging to observe; yet, gas is fundamentally important to determine temperature 
and density gradients in the disk, to drive a chemistry forming important molecules, and to control the
dynamics of the dust itself. Growing, massive, gas-rich planets will accrete their extended atmospheres 
from this gas reservoir. Knowing the gas content and the conditions of the gas phase is therefore fundamental 
to develop models of planet formation, a deeper  understanding of the early evolution of planetary atmospheres,
and eventually of the chemistry on planets. However, gas disks are difficult to observe because high spatial 
resolution spectroscopy in the mm and sub-mm range is still challenging, and optically thin molecular lines 
in the mid-infrared are often faint; promising, pioneering studies have been performed 
\citep{dutrey97, simon00, lahuis06, lahuis07, bitner07}, including observations of CO and H$_2$ 
in the inner, ``terrestrial'' planet-forming disk region \citep{herczeg02, najita03, blake04}
and also detections of organic molecules and water at larger radii \citep{lahuis06, carr08, salyk08}. 

Recent studies, both theoretical \citep{glassgold04, jonkheid04, kamp04, nomura05, glassgold07} and 
observational \citep{weintraub00, herczeg02} have suggested that short-wavelength radiation (ultraviolet 
[UV], extreme ultraviolet [EUV], and X-rays) from the central star significantly affects
circumstellar disks in several ways.  High-energy photons are capable of ionizing the upper layer of
circumstellar disks at  a level orders of magnitude in excess of what cosmic rays could achieve \citep{igea99,
glassgold04, ilgner06}. Even weak ionization of the surface layer will, in combination with magnetic fields,
induce the magnetorotational instability \citep{balbus91}, leading to accretion from the surface layer while
the deeper zones of the disk may remain shielded and thus provide ideal environments for planet formation.

Detection of the environmental impact of X-rays and EUV photons is, however, challenging. Some direct evidence 
for the relevant processes is provided by X-rays themselves; fluorescence of cold iron as seen in the X-ray range
at 6.4~keV has been interpreted as being due to irradiation of the disk surface by hard X-ray photons from
the stellar magnetosphere \citep{imanishi01, tsujimoto05, favata05}. Photoelectric absorption of stellar X-rays
by disk material can be directly measured in particular in cases for which the disk is seen at high inclination
angles \citep{kastner05}. 

The same interactions also heat the disk surface layers to several thousand K, generating
a ``disk chromosphere'' that decouples from the dust component. 
Strong UV continuum and fluorescence \citep{bergin04, herczeg02, herczeg06}
and pure rotational and rovibrational IR transitions of H$_2$ \citep{weintraub00, bary03, bitner07} require 
high ($\sim$1000~K) disk gas temperatures. This warm disk gas may also have been detected through 
CO emission \citep{najita03, blake04, brittain07} and in fine-structure
transitions of [Ne\,{\sc ii}] after ionization by EUV/X-ray photons, as further 
discussed below. As a most important by-product of ionization and heating, chemistry is driven across 
temperature gradients in the disks, especially in UV shielded regions \citep{aikawa99}. 
Furthermore, disk photoevaporation due to X-ray or EUV heating of the inner disk has attracted increasing 
attention \citep{alexander04, ercolano08, gorti09}.

Short-wavelength disk irradiation is thus of central interest for our understanding of
the ionization of circumstellar disks, their heating and chemical processing, disk instabilities, and 
photoevaporation and therefore the long-term evolution of disks. All these mechanisms obviously affect 
the process of planet formation.

\section{The Ne\,{\sc ii} 12.81$\mu$m diagnostic}

\citet{glassgold07} proposed that the mid-infrared [Ne\,{\sc ii}] fine-structure transition 
at 12.81$\mu$m is a tracer of {\it warm} gas requiring X-ray irradiation
of the disk. Because the first ionization potential of Ne is high (21.6~eV), its photoionization
indeed requires EUV or X-ray photons. In the X-ray ionization model of \citet{glassgold07}
Ne is ionized by K-shell absorption, requiring photons with energies of at least 0.9~keV.
As the same X-rays also heat the upper layers of the gas disk to several 1000~K \citep{glassgold04}, [Ne\,{\sc ii}] 
fine-structure transitions with excitation temperatures of $\approx 1000$~K are produced over a scale 
height of warm gas of $10^{19}-10^{20}$~cm$^{-2}$. An alternative model (\citealt{hollenbach09},
see also \citealt{gorti08}) proposes ionization of the disk surface by 
EUV radiation, producing an H\,{\sc ii}-like highly-ionized region in which the [Ne\,{\sc ii}] line 
is formed. The [Ne\,{\sc ii}] transition thus appears to be an ideal tracer both of disk gas  
and of the environmental impact of high-energy stellar radiation. 

[Ne\,{\sc ii}] emissivities were also calculated for shocks in molecular clouds, 
e.g. such as those forming when a protostellar jet rams into the surrounding molecular gas (e.g., \citealt{hollenbach89}). 
The Infrared Space Observatory (ISO) indeed detected the [Ne\,{\sc ii}] line feature in the T Tau 
system \citep{ancker99}; 
the authors attributed this line to shocks in the outflows of the T Tau system. The [Ne\,{\sc ii}] 
line has also been seen in Herbig-Haro objects \citep{neufeld06}.

The advent of the {\it Spitzer Space Telescope} ({\it Spitzer} henceforth; \citealt{werner04}) has renewed interest
in the [Ne\,{\sc ii}] feature from pre-main sequence stars as theoretical calculations of X-ray  irradiated
circumstellar disks \citep{glassgold07} predicted easy detection by the Infrared 
Spectrometer (IRS; \citealt{houck04}). Follow-up calculations have confirmed and deepened these
initial predictions \citep{meijerink08}, also including EUV irradiation \citep{hollenbach09}. 
First successful detections of the [Ne\,{\sc ii}] line by the IRS were reported by
\citet{pascucci07}, \citet{espaillat07}, \citet{lahuis07}, and \citet{ratzka07} from a variety of pre-main sequence objects with 
circumstellar disks and partly also with protostellar envelopes. Initial attempts were made
to relate the observed [Ne\,{\sc ii}]  fluxes to stellar or disk properties, but the mostly small 
samples yielded ambiguous results. \citet{pascucci07} claimed a correlation between the luminosity in the [\rm Ne\,{\sc ii}]
line, $L_{\rm [Ne~II]}$, and the stellar X-ray luminosity, $L_{\rm X}$, but the sample contained only
four detections distributed in $L_{\rm X}$  over a mere 0.2~dex, which corresponds to the usual range of variability of 
stellar coronal X-ray emission. A correlation with the stellar mass accretion rate, $\dot{M}_{\rm acc}$, 
was not found. Conversely, \citet{espaillat07} suggested  $L_{\rm [Ne~II]}$ to be correlated with 
$\dot{M}_{\rm acc}$ but not with $L_{\rm X}$, but again, the data sample and the dynamic range in $L_{\rm X}$ 
were small ($\approx 0.8$~dex). Finally, \citet{lahuis07} considered [Ne\,{\sc ii}] line production both as a 
consequence of X-ray or EUV irradiation of disks and of shock formation on the disks themselves. They noticed 
that the  measured $L_{\rm [Ne~II]}$ correspond well to predictions made by \citet{glassgold07}. 

Ground-based observations of the [Ne\,{\sc ii}] line allow for much higher spectral resolving power, uncovering
the kinematics of the emitting regions. \citet{herczeg07} observed a relatively narrow profile in the near-pole-on
system TW Hya, interpreting the emission as coming either from the disk surface (i.e., from a region where the gas 
is gravitationally bound), or a slow photoevaporative flow from a disk region that allows for escaping gas flows
given a sufficiently high gas temperature.
[Ne\,{\sc ii}] emission from such flows was indeed modeled by \citet{alexander08}, their results being consistent with
the observations. Additional support came from high spectral resolution observations 
of transition disks; three objects of this class showed line profiles and blue-shifts consistent with those predicted  
from photoevaporative flows \citep{pascucci09}.
Further ground-based observations of AA Tau and GM Aur with the TEXES instrument clearly show that most of their [Ne\,{\sc ii}]
emission is consistent with a disk origin although the lower fluxes, when compared with {\it Spitzer} results,
suggest that there is also an extended component (not recovered by the narrow slit of TEXES). Alternatively,
the sources may be time-variable, or the line is spectrally unusually broad \citep{najita09}.

A rather unexpected twist came with the observation of the T Tauri system with the VLT at high spectral resolution
\citep{boekel09}. Here, the [Ne\,{\sc ii}] emission region was clearly extended (by several arcseconds) and showed line broadening 
and line shifts (by up to 126~km~s$^{-1}$) compatible with the jets of the T Tauri system, previously observed in a similar 
fashion in [S\,{\sc ii}] and [O\,{\sc i}] by \citet{boehm94}. The interpretation favored shock-induced 
[Ne\,{\sc ii}] formation, but X-ray irradiation by the star and consequent ionization of the jet material remains 
a viable option \citep{boekel09}. This scenario (stellar X-rays ionizing the jet, and [Ne\,{\sc ii}] emission forming in the 
jet itself) has recently been suggested on theoretical grounds as well \citep{shang10}.

The variety of possible formation scenarios of the [Ne\,{\sc ii}] line and the large range of stellar and disk
properties, including the presence of jets and outflows in some of the systems, calls for a deeper investigation requiring
a much larger sample. To this end, we have started a comprehensive study of [Ne\,{\sc ii}] emission from classical
T Tauri stars (CTTS) in particular in the context of X-ray emission and measured mass accretion rates. Our goal has been to revisit
recent work that was based mostly on small samples or observations of individual objects, and to analyze and study the combined
sample of objects in a systematic  way. We have re-analyzed many of the previously reported [Ne\,{\sc ii}]
data (mostly from {\it Spitzer}) coherently, but we have also added crucial new observations from our dedicated observing
programs. The present work also for the first time presents a uniform, archival study of all available X-ray data from the {\it XMM-Newton}
and {\it Chandra} observatories for these sources, complemented with data from new observing programs. The 
increased sensitivity and spectral resolving power of these X-ray devices permit a much better characterization of the
stellar X-ray radiation than hitherto possible with data from, e.g., ROSAT  as used in earlier [Ne\,{\sc ii}] studies
\citep{lahuis07, pascucci07}. We will further use ancillary data collected from the published literature, such as mass 
accretion rates or mass outflow rates. We also refer to forthcoming work by \citet{baldovin10} in which a sample of
gas lines (of H$_2$, [Ne\,{\sc ii}], [Ne\,{\sc iii}], [Fe\,{\sc ii}], [S\,{\sc iii}]) in the IRS spectal range
is discussed but for a more confined sample of pre-main sequence stars in Taurus; few significant correlations 
are reported there, the most likely one again supporting an origin of [Ne\,{\sc ii}] emission in outflows.

Initial results of our work are described
in \citet{guedel09a}. In short, indications of a correlation of $L_{\rm [Ne~II]}$ with both $L_{\rm X}$  and
$\dot{M}_{\rm acc}$ were found, but strong scatter dominates these correlations. However, it was also found that 
{\it sources with jets} show consistently  higher $L_{\rm [Ne~II]}$, and that $L_{\rm [Ne~II]}$ seems to correlate 
with wind or outflow properties. The purpose of the present paper is to present our entire data set and additional 
correlation studies, and to coherently discuss these results.

\section{Stellar sample}

Given the presently favored models for [Ne\,{\sc ii}] line emission, we selected targets for our study that
have well-observed disks and may also be engines of jets an outflows, but we did not include Class I objects in 
which a number of additional circumstellar regions may be relevant for [Ne\,{\sc ii}] emission, such as shocks on disks
produced by material accreting from the envelope, the irradiated envelopes themselves, shocks between the jets or
outflows and the envelopes, etc. Further, strong extinction and photoelectric absorption make Class I objects difficult
for study. Sufficiently strong extinction may suppress some infrared emission from the regions close to the star selectively.
Also, only a moderate fraction of Class I sources is accessible by modern X-ray satellites, and the detected X-ray emission
is, due to photoelectric absorption, relatively hard (photon energies typically $>$2~keV). The bulk part of the X-ray
emission therefore remains undetected, and an unbiased reconstruction of the underlying (intrinsic) X-ray  emission relies
on assumptions. For a  study predominantly based on [Ne\,{\sc ii}] detections of Class I sources, see
\citet{flaccomio09}.

Our targets were therefore required
to be essentially Class II objects or (accreting) CTTS. More massive Herbig stars were
not considered given their considerable UV radiation fields and possibly very different X-ray source properties 
\citep{telleschi07a}.  On the other hand, CTTS ejecting jets were intentionally included because jets may play a crucial
role in strongly accreting CTTS, and they may be directly linked to the accretion process itself.
Including such objects will therefore allow us to investigate to what extent jets matter for the observed
[Ne\,{\sc ii}] emission, and perhaps to identify a subset of objects showing a baseline [Ne\,{\sc ii}] flux unaffected
by jets. To study this latter possibility further, a few targets revealing signatures
of transition disks, i.e., disks with inner holes, have been included.  Transition disks may also be 
important to discriminate between X-ray and EUV-related [Ne\,{\sc ii}] emission models as the potentially low 
gas content in the inner disk may make this region transparent to direct EUV radiation 
(see, e.g., numerical models by \citealt{alexander06}).

The data selection is primarily driven by the availability of [Ne\,{\sc ii}] observations (detections or upper limits).
Our sample is therefore mostly drawn from observations available in the {\it Spitzer} IRS data archive.
The largest part of our target list originates from the \citet{lahuis07} survey (based on the 
{\it Spitzer} {\it Cores to Disk} {\it [c2d]} legacy program; \citealt{evans03}). This survey focuses
on the NGC~1333, Chamaeleon, Lupus, Rho Oph, and Serpens star forming regions, all with characteristic ages of a few Myr. 
Compared to the preliminary presentation in \citet{guedel09a}, we have reduced these data again using a new, improved 
software version with more careful background subtraction and treatment of potential blends, resulting in 
many additional detections not used in the initial report. 

To this sample, we added several targets from our {\it Spitzer general observer} programs (PI J. Carr).
This sample in particular includes objects from the Taurus star forming region, and some targets from Chamaeleon
and Rho Oph. Further [Ne\,{\sc ii}] fluxes or upper 
limits thereof were adopted from the published literature, in particular for RX~J1111.7-7620, PZ99~J161411, 
RX~J1842.9-3542, and RX~J1852.3-3700 \citep{pascucci07}, observed as part of the {\it Spitzer Formation and Evolution
of Planetary Systems (FEPS)} legacy program \citep{meyer04},  DP Tau observed with the {\it MICHELLE} spectrograph at
Gemini North \citep{herczeg07}, and T Tau N and S observed with the {\it VISIR} spectrograph at
the Very Large Telescope (VLT) \citep{boekel09}. These references describe the respective data reduction in detail.

Our targets and their properties are listed in Tables~\ref{table1}--\ref{table4} (the first ten entries are
displayed; the complete tables are available in the electronic version of this paper). Table~\ref{table1} lists
the adopted stellar names, the same as those used in the original literature reporting portions of our [Ne\,{\sc ii}]
sample; some common alternative names are also given. The table further gives J2000.0 coordinates 
and adopted distances; all entries are ordered in increasing RA. The targets are arranged identically in the 
subsequent three tables. Table~\ref{table2}  lists observing parameters, namely the Spitzer {\it Astronomical 
Observation Request} (AOR; unless observed
by another observatory, as indicated), the X-ray observation ID (referring to {\it XMM-Newton} if not 
otherwise noted), and  the X-ray observation start and stop times together with the total exposure. We note 
that the full exposure time was not normally used for {\it XMM-Newton} data analysis as time intervals with high 
particle background were eliminated to achieve an optimum signal-to-noise ratio (see 
\citealt{guedel07a} for details). Table~\ref{table3} lists our primary results, namely the observed
fluxes in the [Ne\,{\sc ii}] line and the photoelectrically {\it attenuated} (absorbed, observed) and the modeled 
{\it intrinsic}, unabsorbed 
X-ray fluxes in the 0.3--10~keV range (i.e., the flux measured at Earth if absorption were absent).  These fluxes are
complemented by the respective luminosities using the distances from Table~\ref{table1}. We also give  
$N_{\rm H}$ along the line of sight to the star, as derived from the X-ray spectral fits (see below).  All values 
are given to two significant digits, where uncertainties typically affect the second digit.

Finally, Table~\ref{table4} provides selected ancillary data extracted from the published literature.
The parameters listed here are described below. Column 2 gives the mass accretion rate onto the star, 
$\dot{M}_{\rm acc}$. Mass accretion rate determination is based on various methods (e.g., spectrophotometry of veiling of 
absorption lines due to blue continuum, or also photometry of U band excess in \citealt{hartmann98}, the two methods
producing compatible results; similar methods based on UV/optical accretion excess emission also for other references
as summarized in \citealt{najita07}; \citealt{natta06} used Pa$\beta$ and Br$\gamma$ lines and compared with 
other methods). 
As $\dot{M}_{\rm acc}$ varies among different authors by up to an oder of magnitude or so, we adopted the homogenized accretion 
rates listed by \citet{najita07} or used, as far as possible, values from publications compatible with this compilation
(in particular \citealt{hartmann98}).
Column 3 gives the equivalent width of the H$\alpha$ line, EW(H$\alpha$), and column 4 the equivalent width of the 
[O\,{\sc i}]~$\lambda$6300 line (for total flux, index $t$). This line is suspected to originate in jets and outflows, 
although this interpretation is based on the ``high-velocity'' component sometimes seen
in the line \citep{hartigan95}. If such a component has been separately measured, we list in the next column
(col. 5) the equivalent width of the high-velocity component of the [O\,{\sc i}] line (index $f$ for ``fast''); we then
give, in column 6, the logarithm of the [O\,{\sc i}] luminosity normalized with the solar luminosity, for the high-velocity component ($f$). 
To convert EW([O\,{\sc i}]) to $L_{\rm [O\,I]}$, we used the method described by \citet{hartigan95} based on the R-band 
magnitude and the visual extinction; these parameters were taken from the literature or from the SIMBAD database and the 
2MASS catalog \citep{skrutskie06}. Likewise, in column 7 we give the luminosity corresponding to the total  flux ($t$)
in the [O\,{\sc i}] line, derived from the corresponding EW as above; for some objects, a ``wind mass-loss
rate'' is available in the literature (given in column 8), as derived for example from the [O\,{\sc i}]$_f$ line flux. 
We also indicate, in column 9, objects for which explicit evidence for jets or outflows has been reported in the literature (``J'') and
objects classified as having transition disks (``T''). Finally, the last
column (col. 10) in Table~\ref{table4} lists references from which these data were obtained (sequentially for the 
columns in the table), or from which auxiliary parameters such as R-band magnitudes and $A_{\rm V}$ (for the EW-$L$ 
conversion) were adopted. Specifically, most values for $\dot{M}_{\rm acc}$ are from \citet{najita07} and \citet{hartmann98}.
EW(H$\alpha$) are mostly taken from \citet{cohen79} and \citet{hughes94} (the latter for Lupus). For Taurus objects, 
parameters were extracted 
from \citet{guedel07a} where further references are listed. Information on [O\,{\sc i}] equivalent widths and luminosities
are mostly from \citet{cohen79}, \citet{hamann94}, \citet{hartigan95}, and \citet{hirth97}. Most of 
the mass loss rates, $\dot{M}_{\rm loss}$, are from \citet{hartigan95}. Although some of the values listed in 
Table~\ref{table4} may be given to excessive precision, we prefer leaving the values extracted from the 
published literature unaltered.

A few notes on individual targets follow: BYB~35 was observed in X-rays but remained
undetected. As an approximate  extinction is known  ($A_{\rm J} = 5.9$~mag, \citealt{gomez03}),   
we estimated the expected $N_{\rm H}$ based on the standard interstellar gas-to-dust mass ratio ($\approx 100$)
and dust properties, using $N_{\rm H} \approx 2\times 10^{21}A_{\rm V}$~cm$^{-2}$  and $A_{\rm V} = 3.6\times
A_{\rm J}$  (e.g., \citealt{vuong03}). We then adopted a standard X-ray emission model as found for other CTTS 
in Taurus (see \citealt{guedel07a} for details) to estimate the flux upper limits based on the background count rate 
in the vicinity of the expected stellar position.

The two targets SSTc2d~J182928.2+002257 and SSTc2d~J182909.8+003446 were also observed in X-rays but both 
remained undetected. Given the poorly known properties of these objects and the absence of reliable 
extinction values to estimate expected $N_{\rm H}$ values, no reliable upper limits to the X-ray fluxes 
and luminosities could be calculated.

T Tau S is occasionally defined as a protostar. Although Class I sources are not studied here (see \citealt{flaccomio09} for 
such objects), we do include 
T Tau S, itself a binary, together with T Tau N.  
The status of T Tau S is not entirely clear - much of the observed extinction/absorption may in fact be due to a thick, 
near-edge-on disk \citep{solf99, duchene05}. Also, T Tau S is the best studied [Ne\,{\sc ii}] jet source, providing important evidence for
the role of outflows in the production of [Ne\,{\sc ii}]  emission.
T Tau S has not clearly been detected in X-rays owing to high photoelectric absorption, although 
marginal excess flux at its position may be present \citep{guedel07c}. 
However, the mass of the more massive component T~Tau Sa is very nearly the same as the mass of T Tau N,
namely $\approx 2M_{\odot}$ \citep{koehler08}. Because X-ray emission of CTTS is correlated with stellar mass 
\citep{telleschi07b}, we assign the same $L_{\rm X}$ to T Tau Sa as found for T Tau N. This value may be uncertain
by a factor of a few but given the outstanding [Ne\,{\sc ii}] properties of this object, in particular its
high [Ne\,{\sc ii}] luminosity \citep{boekel09}, this uncertainty will not critically influence our results. The X-ray
luminosity is in
rough agreement with an estimated X-ray luminosity corresponding to the marginal excess flux seen in the 
{\it Chandra} HRC image  \citep{guedel07c}.

DG~Tau, DP~Tau, and HN~Tau show peculiar X-ray spectra with two components subject to different absorbing gas column
densities \citep{guedel07b, guedel09b}. We considered only the hard, coronal component for the X-ray flux, while the
soft component is probably associated with the jets.
In Sz~102 (= TH 28, or ``Krautter's Star''), the entire observed X-ray flux may be related to jets \citep{guedel09b}. 
Its X-ray spectrum is very soft, while the expected near-edge-on geometry should absorb essentially all
stellar X-rays or transmit only the hardest portion of the spectrum. We will therefore not consider this star for
statistical studies involving $L_{\rm X}$.

Table~\ref{table5} summarizes sample statistics. In total, our sample contains 92 objects, for all of
which we derived [Ne\,{\sc ii}] fluxes or upper limits or found corresponding information in the literature
(58 detections and 34 upper limits). X-ray information is available for 67 of these objects, 64 of which
were detected. Both [Ne\,{\sc ii}] and X-ray detections are available for 40 objects. Obviously, ancillary
data are far from complete for our sample, and therefore smaller subsets had to be used for specific 
correlation studies.

\begin{table*}
\caption{Targets [first ten entries]}
\begin{tabular}{llrrr}
\hline
\hline
Star                    & Alternative                                                                                     &RA(J2000.0)& dec(J2000.0)              & dist.	\\					 
                        & names                                                                    			  & (h m s)  & (deg $\arcmin$ $\arcsec$)  & (pc)        \\
\hline
RNO~15                  & HBC~339, CoKu~NGC~1333/2                                                                        & 03 27 47.7  &  30 12 04.3    &   250  \\				      
LkH$\alpha$~270         & HBC~12, NGC~1333~IRS~2                                                   			  & 03 29 17.7  &  31 22 45.1 	 &   250  \\
LkH$\alpha$~271         & HBC~13                                                                   			  & 03 29 21.9  &  31 15 36.4 	 &   250  \\
LkH$\alpha$~326         & HBC~14                                                                   			  & 03 30 44.0  &  30 32 46.7 	 &   250  \\
LkH$\alpha$~327         & HBC~15, IRAS~03304+3100                                                  			  & 03 33 30.4  &  31 10 50.5 	 &   250  \\
LkH$\alpha$~330         & HBC~20, IRAS~F03426+3214                                                 			  & 03 45 48.3  &  32 24 11.9 	 &   250  \\
IRAS 03446+3254         & IRS~4                                                                    			  & 03 47 47.1  &  33 04 03.4 	 &   250  \\
BP~Tau                  & HBC~32, HD~281934, IRAS~04161+2859                                       			  & 04 19 15.8  &  29 06 26.9 	 &   140  \\
FM~Tau                  & HBC~23, Haro~6-1                                                                                & 04 14 13.6  &  28 12 49.2    &   140  \\									      
T~Tau N                 & HBC~35, HD 284419, IRAS~04190+1924                                       			  & 04 21 59.4  &  19 32 06.4 	 &   140  \\
\hline                  
\end{tabular}           
\label{table1}            
\normalsize             
\end{table*}            
                        
\begin{table*}
\caption{Mid-IR and X-ray observations [first ten entries]}
\begin{tabular}{lrrrrr}
\hline
\hline
Star                    & {\it Spitzer}& X-ray       & X-ray start time$^c$  &  X-ray stop time$^c$  & Total X-ray    \\
                        & AOR$^a$      & ObsID$^b$   & y-m-d h:m:s           &  y-m-d h:m:s	     & exposure$^c$ (s) 	 	   \\
\hline
RNO~15                  & 5633280      & 0503670101  & 2007-07-31\ 04:55:25  & 2007-07-31\ 15:32:31  & 33135	     \\ 	      
LkH$\alpha$~270         & 5634048      & 0065820101  & 2002-02-27\ 22:48:25  & 2002-02-28\ 12:41:45  & 44808	    \\ 
LkH$\alpha$~271         & 11827968     & 0065820101  & 2002-02-27\ 22:48:25  & 2002-02-28\ 12:41:45  & 44808	     \\       
LkH$\alpha$~326         & 5634304      & ...	     & ...		     & ...		     & ...	    \\       
LkH$\alpha$~327         & 5634560      & ...	     & ...		     & ...		     & ...	    \\       
LkH$\alpha$~330         & 5634816      & ...	     & ...		     & ...		     & ...	    \\       
IRAS 03446+3254         & 5635072      & ...	     & ...		     & ...		     & ...	    \\         
BP~Tau                  & 14548224     & 0200370101  & 2004-08-15\ 06:36:51  & 2004-08-16\ 18:42:57  & 116334       \\
FM~Tau                  & 15119872     & 0203542001  & 2004-09-12\ 07:21:01  & 2004-09-12\ 15:47:38  & 26760        \\	
T~Tau N                 & (VLT)	       & 0301500101  & 2005-08-15\ 14:14:33  & 2005-08-16\ 12:55:22  & 65810	    \\
\hline                  
\end{tabular} \\          
Notes:\\
$^a$ {\it Spitzer} AOR = Astronomical Observation Request; for further details on observing setup, see references in Table~\ref{table3}. \\
$^b$ {\it XMM-Newton} observation identification number if not otherwise noted; {\it Chandra} ID if `(CXO)' added. \\
$^c$ Exposure start and stop times for the {\it XMM-Newton} EPIC pn camera if not otherwise noted; if (M1), (M2), (M) added, then MOS1, MOS2, or both
     MOS were used for analysis, and exposure times refer to those detectors; (+M12) indicates that MOS detectors were
     used additionally to PN for spectral analysis. Exposure times in last column are livetimes for CCD\#1 of detector and are only 
     indicative as intervals with high background were selectively and additionally flagged.\\ 
\label{table2}            
\normalsize             
\end{table*}            

\begin{table*}[t!]
\caption{Fluxes and luminosities [first ten entries]}
\begin{tabular}{lrrrrrrrl}
\hline
\hline
Star                     & $f_{\rm [Ne\,II]}^a$		         & $L_{\rm [Ne\,II]}$	   &$f_{\rm X, 0.3-10, abs}$&$L_{\rm X, 0.3-10, abs}$&$f_{\rm X, 0.3-10, unabs}$&$L_{\rm X, 0.3-10, unabs}$& $N_{\rm H}^b$  & Refs.$^c$\\ 
                         &  (erg~cm$^{-2}$~s$^{-1}$)		 &      (erg~s$^{-1}$)	   &(erg~cm$^{-2}$~s$^{-1}$)& (erg~s$^{-1}$)         &(erg~cm$^{-2}$~s$^{-1}$)  & (erg~s$^{-1}$)           &   ($10^{22}$)  &  \\
\hline
RNO~15                   & $<2.6(0.85)\phantom{ }\times 10^{-14} $ & $<1.9\times 10^{29}$  &  $3.7\times 10^{-13}$  & $2.8\times 10^{30}$  &  $1.1\times 10^{-12}$ &  $8.0\times 10^{30}$  & 4.6       & 1,X	 \\  
LkH$\alpha$~270          & $ 1.3(0.35)\phantom{ }\times 10^{-14} $ & $ 9.7\times 10^{28}$  &  $3.2\times 10^{-13}$  & $2.4\times 10^{30}$  &  $1.5\times 10^{-12}$ &  $1.1\times 10^{31}$  & 0.78      & 1,X	     \\  
LkH$\alpha$~271          & $<3.6(1.2)\phantom{0}\times 10^{-15}  $ & $<2.7\times 10^{28}$  &  $3.8\times 10^{-14}$  & $2.8\times 10^{29}$  &  $1.1\times 10^{-13}$ &  $7.9\times 10^{29}$  & 3.8       & 1,X	     \\  
LkH$\alpha$~326          & $ 3.1(1.4)\phantom{0}\times 10^{-15}  $ & $ 2.3\times 10^{28}$  &  ...		    & ...                  &  ...	           &  ...                  & ...       & 1	     \\  
LkH$\alpha$~327          & $ 7.8(3.8)\phantom{0}\times 10^{-15}  $ & $ 5.8\times 10^{28}$  &  ...		    & ...                  &  ...	           &  ...                  & ...       & 1	     \\  
LkH$\alpha$~330          & $ 3.8(1.9)\phantom{0}\times 10^{-15}  $ & $ 2.8\times 10^{28}$  &  ...		    & ...                  &  ...	           &  ...                  & ...       & 1	     \\  
IRAS 03446+3254          & $ 4.3(0.99)\phantom{ }\times 10^{-15} $ & $ 3.2\times 10^{28}$  &  ...		    & ...                  &  ...	           &  ...                  &   ...     & 1,X	     \\  
BP~Tau                   & $ 2.9(0.4)\phantom{0}\times 10^{-15}  $ & $ 6.8\times 10^{27}$  &  $4.3\times 10^{-13}$  & $1.0\times 10^{30}$  &  $6.1\times 10^{-13}$ &  $1.4\times 10^{30}$  & 0.09      & 3,X	     \\  
FM~Tau                   & $<1.1(0.37)\phantom{ }\times 10^{-14} $ & $<2.6\times 10^{28}$  &  $1.5\times 10^{-13}$  & $3.5\times 10^{29}$  &  $2.2\times 10^{-13}$ &  $5.1\times 10^{29}$  & 0.15      & 3,X	     \\  
T~Tau N                  & $ 2.0(0.4)\phantom{0}\times 10^{-13}  $ & $ 4.7\times 10^{29}$  &  $1.9\times 10^{-12}$  & $4.4\times 10^{30}$  &  $4.0\times 10^{-12}$ &  $9.4\times 10^{30}$  & 0.34      & 5,X	     \\  
\noalign{\smallskip}\hline
\end{tabular}\\
Notes:\\
$^a$ Errors are given in parentheses.   \\
$^b$ In units of $10^{22}$~cm$^{-2}$.   \\
$^c$ References: 1: New analysis of Spitzer c2d sample, for original analysis see \citet{lahuis07}; 2: \citet{pascucci07}; 3: archival GTO/GO observations, see Najita \& Carr, in prep.;
                 4: \citet{herczeg07}; 5: \citet{boekel09}; 6: \citet{espaillat07}; C: from {\it Chandra} archive; X: from {\it XMM-Newton} archive.   \\
\label{table3}            
\end{table*}

\begin{table*}[t!]
\caption{Additional parameters [first ten entries]}
\begin{tabular}{lrrrrrrrrr}
\hline
\hline
Star                     & $\log \dot{M}_{\rm acc}$ &  EW(H$\alpha$)  & EW([O\,{\sc i}]$_f$) &   EW([O\,{\sc i}]$_t$) & $\log L_{{\rm [OI]},_f}$  & $\log L_{{\rm [OI]},t}$  & $\log \dot{M}_{\rm loss}$    & jet? & Refs.$^b$   \\ 
                         & ($M_{\odot}$~yr$^{-1}$)  &      (\AA)      & (\AA)                &   (\AA)                & ($L_{\odot}$)             & ($L_{\odot}$)          & ($M_{\odot}$~yr$^{-1}$)   & TD?$^a$  &   \\
\hline
RNO~15                   &  ...   & ...    &     ...   &  ...   &   ...  &    ...  &    ...  &   ...  &  ...\\  
LkH$\alpha$~270          &  ...   &   30.9 &     ...   &  ...   &   ...  &    ...  &    ...  &   ...  &  6 \\  
LkH$\alpha$~271          &  ...   & 185.7  &      ...  &  ...   &   ...  &    ...  &    ...  &   ...  &  6\\  
LkH$\alpha$~326          &  ...   &   52.7 &     ...   & 0.8    &   ...  &    ...  &    ...  &   ...  &  6 \\  
LkH$\alpha$~327          &  ...   &   51.0 &     ...   & 0.9    &   ...  &    ...  &    ...  &   ...  &  6 \\  
LkH$\alpha$~330          &  ...   &   20.3 &     ...   &  ...   &   ...  &    ...  &    ...  &   T    &  6,31 \\  
IRAS 03446+3254          &  ...   &  ...   &     ...   & ...    &   ...  &    ...  &    ...  &   ...  &  ... \\  
BP~Tau                   & -7.54  &  40-92 &    0.07   &   0.26 &  -5.71 & -4.75   & $<$-9.7 &   ...  &  13,X,12 \\  
FM~Tau                   & -8.45  &  62-71 &     0.045 & 0.48   &  -6.37 &  -5.19  & $<$-10.6&   ...  &  21,35,6,12       \\  
T~Tau N                  & -7.12  &  38-60 &     ...   &  2.0   &   ...  &  -2.78  &  -6.96  &   J    &  21,6,26,4,(27) \\  
\noalign{\smallskip}\hline
\end{tabular}\\
Notes:\\
$^a$ T = transitional disk (either ``anemic'' or ``cold'' disk); J = jet-driving object.\\
$^b$ References:
    1 \citet{alcala08}; 
    2 \citet{bacciotti99};
    3 \citet{bally06};
    4 \citet{boehm94};
    5 \citet{briceno02};
    6 \citet{cohen79};
    7 \citet{comeron03}; 
    8 \citet{espaillat07};
    9 \citet{gauvin92};
    10 \citet{hamann94};
    11 \citet{hartigan93};
    12 \citet{hartigan95};
    13 \citet{hartmann98};
    14 \citet{herczeg04};
    15 \citet{herczeg05};
    16 \citet{herczeg07};
    17 \citet{hirth97};
    18 \citet{hughes94};
    19 \citet{krautter97};
    20 \citet{luhman04};
    21 \citet{najita07};
    22 \citet{pascucci07};
    23 \citet{rydgren80};
    24 \citet{seperuelo08};
    25 \citet{takami03};
    26 \citet{white04};
    27 \citet{kenyon95};
    28 \citet{mundt98};
    29 \citet{solf93};
    30 \citet{cox05};
    31 \citet{brown07};
    32 \citet{calvet02};
    33 \citet{calvet05};
    34 \citet{forrest04};
    35 \citet{white01};
    36 \citet{dougados00};
    37 \citet{hartigan04};
    38 \citet{najita08};
    39 \citet{natta06};
    40 \citet{webb99};
    41 \citet{alcala93}.
    M: 2MASS; 
    S: SIMBAD; 
    X: XEST survey, see \citet{guedel07a} and references therein.
    References in parentheses are for complementary data, e.g., $A_{\rm V}$ or R magnitude, used to calculate $\log L_{\rm [OI]}$.  \\
\label{table4}            
\end{table*}

\begin{table}[t!]
\caption{Sample statistics}
\begin{tabular}{lr}
\hline
\hline
                                          & Number of targets \\
\hline
Sample                                    &  92\phantom{$^a$} \\
\hline
{\rm [Ne\,{\sc ii}] observations}         &  92\phantom{$^a$} \\
{\rm [Ne\,{\sc ii}] detections}           &  58\phantom{$^a$} \\
{\rm [Ne\,{\sc ii}] non-detections}       &  34\phantom{$^a$} \\
\hline
X-ray observations                        &  67\phantom{$^a$} \\
X-ray detections                          &  64$^a$ \\
X-ray non-detections                      &   3\phantom{$^a$} \\
\hline
{\rm [Ne\,{\sc ii}]} \& X-ray  observations                 &  67\phantom{$^a$} \\
{\rm [Ne\,{\sc ii}]} detection \& X-ray detection           &  40$^a$ \\
{\rm [Ne\,{\sc ii}]} non-detection \&  X-ray detection      &  24\phantom{$^a$} \\
{\rm [Ne\,{\sc ii}]} detection \& X-ray non-detection       &   1\phantom{$^a$} \\
{\rm [Ne\,{\sc ii}]} non-detection \& X-ray non-detection   &   2\phantom{$^a$} \\
\noalign{\smallskip}\hline
\multicolumn{2}{l}{$^a$ T Tau S included in X-ray detections (for intrinsic $L_{\rm X}$)}
\end{tabular} 
\label{table5}              
\end{table}

\section{Data reduction and analysis}

For a summary of the analysis strategies for the largest [Ne\,{\sc ii}] subsample discussed in our paper, see
\citet{lahuis07}. The objects from Spitzer GO program 2030 (AORs 145XXXXX in
Table~\ref{table2}) were all reduced according to the procedure described in \citet{carr08}.

X-ray data are available from different satellite observatories. We confined our X-ray analysis to
data from the CCD detectors on board {\it XMM-Newton} \citep{jansen01} and the {\it Chandra X-ray Observatory} ({\it Chandra} henceforth; 
\citealt{weisskopf96}). Although {\it ROSAT} observed many of our targets as well, its rather soft bandpass  
(0.1--2 keV) and its very low spectral resolving power ($E/\Delta E \approx 2$) make a reliable modeling of 
relatively faint CTTS subject to considerable  absorption difficult and uncertain. All {\it XMM-Newton}
and {\it Chandra} data were consistently reduced and analyzed. The data reduction  procedures for 
{\it XMM-Newton} data are identical to those described in \citet{guedel07a} for objects in Taurus.
Whenever possible, we extracted the X-ray spectra from the pn-type {\it European Photon Imaging Camera} (EPIC-pn; \citealt{strueder01});
if this camera did not provide useful data (e.g., if the target fell into a CCD gap), we used the two 
spectra extracted from the MOS-type EPIC cameras \citep{turner01}. The few {\it Chandra} spectra were extracted from the 
{\it Advanced CCD Imaging Spectrometer} (ACIS), using the so-called events2 files from the archive. 
Both for {\it XMM-Newton} and {\it Chandra}, counts were extracted from circular areas around the 
source position, and background spectra were defined from nearby, source-free areas on the same CCD chip.

The X-ray spectral interpretation was performed in the XSPEC vers. 11.3.1 software \citep{arnaud96} using 
simple one- or two-component (in exceptional cases, three-component) optically thin, collisional-equilibrium 
plasma models, each component being  defined by its temperature ($T$) and emission measure (EM). The element 
abundances of the plasma were held fixed at values commonly found in pre-main sequence or young active stars 
(see \citealt{guedel07a}). The spectral model was further subject to photoelectric absorption  described by
the absorbing gas (equivalent hydrogen) column density, $N_{\rm H}$. Fit parameters therefore were $T$ and EM for each component, 
and $N_{\rm H}$ in common to all components. We will report only the total X-ray fluxes of our
targets and $N_{\rm H}$, as these are the most important parameters
for theories of [Ne\,{\sc ii}] emission. Fitted EMs roughly scale with $L_{\rm X}$, and temperatures were usually
found in the range typical for T Tauri stars (i.e., a few tenths to a few keV, see \citealt{guedel07a} for
the Taurus objects reported here).

\section{Results}

\subsection{Ne\,{\sc ii} emission from subsamples}

We start the presentation of our results by reviewing the range of evolutionary stages and circumstellar 
environments of our targets. This consideration is motivated by our finding  that jets and
outflows appear to be important contributors to the [Ne\,{\sc ii}] emission from young stars. 

We have identified 14 objects with some evidence of spatially resolved jets or outflows, defining the
subclass of {\it jet sources}. This classification is purely qualitative (e.g., based on imaging in forbidden lines, 
or evidence of Herbig-Haro objects) as no effort was made to quantify mass loss rates, shock speeds, or shock 
excitation in the jets. We do, at this stage, not include objects with indirect evidence for jets, such 
as strong but spatially unresolved [O\,{\sc i}] emission. We will discuss such more quantitative parameters that 
may be related to outflows in a later step.

\begin{figure}[h!]
\includegraphics[angle=0,width=9.cm]{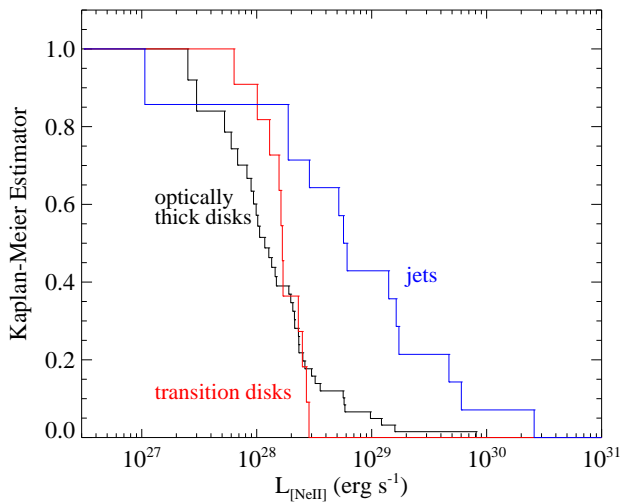}
\caption{Kaplan-Meier estimators for $L_{\rm [Ne\,II]}$ for the three subsamples (optically thick disks without jets,
         transition disks, and jets).}
\label{fig1} 
\end{figure}

Our sample also contains 13 {\it transition disks} which we study separately. Note that we
include different types of transition disks (e.g., \citealt{najita07}). Some of them have low disk
masses and are optically thin (at UV and infrared wavelengths)
throughout the disk. They are sometimes also called ``anemic'' disks
\citep{lada06}. Another class of transition disks are those with
a gap or hole in the dust distribution in the inner disk but with a
massive optically thick outer disk, sometimes also called ``cold'' disks
\citep{brown07}. Several cold disks are now known to have residual
gas present inside the dust gap (e.g., \citealt{pontoppidan08, salyk09}). 
Transition disks rarely show jets, making them a relatively homogeneous group without much [Ne\,{\sc ii}] 
contamination from jets and outflows, although the level of disk clearance will obviously vary among the objects.
CS~Cha is exceptional in this group, showing both a transition disk \citep{espaillat07} and signatures of a jet 
\citep{takami03}. We will 
address this case separately although we will generally include it in the subclass of jets given that jets
may largely dominate [Ne\,{\sc ii}] line emission (see below).
We thus define the remainder of our objects as the class of {\it optically thick disks} without (known) jets (66 objects).

We first study whether the [Ne\,{\sc ii}] production differs between the above three classes. Fig.~\ref{fig1} shows
the Kaplan-Meier estimator for the cumulative distribution of $L_{\rm [Ne\,II]}$, including information from upper 
limits, as calculated in the ASURV statistical software package \citep{lavalley92}.
For {\it optically thick disks}, $L_{\rm [Ne\,II]}$ is broadly distributed between $10^{27}$~erg~s$^{-1}$ and
$10^{30}$~erg~s$^{-1}$, with a median at $1.1\times 10^{28}$~erg~s$^{-1}$.  Interestingly, the distribution seems to be 
more narrowly confined for the smaller sample of {\it transition disks} (excluding CS~Cha), but it shows nearly the same 
median,  $L_{\rm [Ne\,II]} \approx 1.6\times 10^{28}$~erg~s$^{-1}$. The distribution is bounded by the maximum luminosity 
of $\approx 3\times 10^{28}$~erg~s$^{-1}$. The probability that the two distributions are drawn from the same parent 
population is 31\% (using the  Peto-Prentice Generalized Wilcoxon text in ASURV). 

In contrast, the {\it jet sources} show a significantly different 
distribution shifted to nearly $\sim$tenfold higher luminosities, with a median of $6\times 10^{28}$~erg~s$^{-1}$. Here, the
probability that this distribution agrees with the distribution of the optically thick disks is 0.01\%.

We next seek correlations between $L_{\rm [Ne\,II]}$ and stellar or disk parameters. We will employ linear regression
to compute functions of the form  $\log L_{\rm [Ne\,II]} = a + b\log x$, but because many objects show upper limits
to $L_{\rm [Ne\,II]}$, we will use ``survival statistics'' that take these values into account. We use the
parametric estimation maximization (EM) algorithm in ASURV, which implements methods presented by 
\citet{isobe86}. We also use ASURV to compute correlation coefficients for the same samples, specifically using the 
Cox hazard model, Kendall's tau, and Spearman's rho values (where the latter typically requires at least 30
entries to be accurate). A summary of our statistical results is given in Table~\ref{table6}.

\subsection{[Ne\,{\sc ii}] emission and X-rays}
\begin{figure}
\includegraphics[angle=0,width=9.cm]{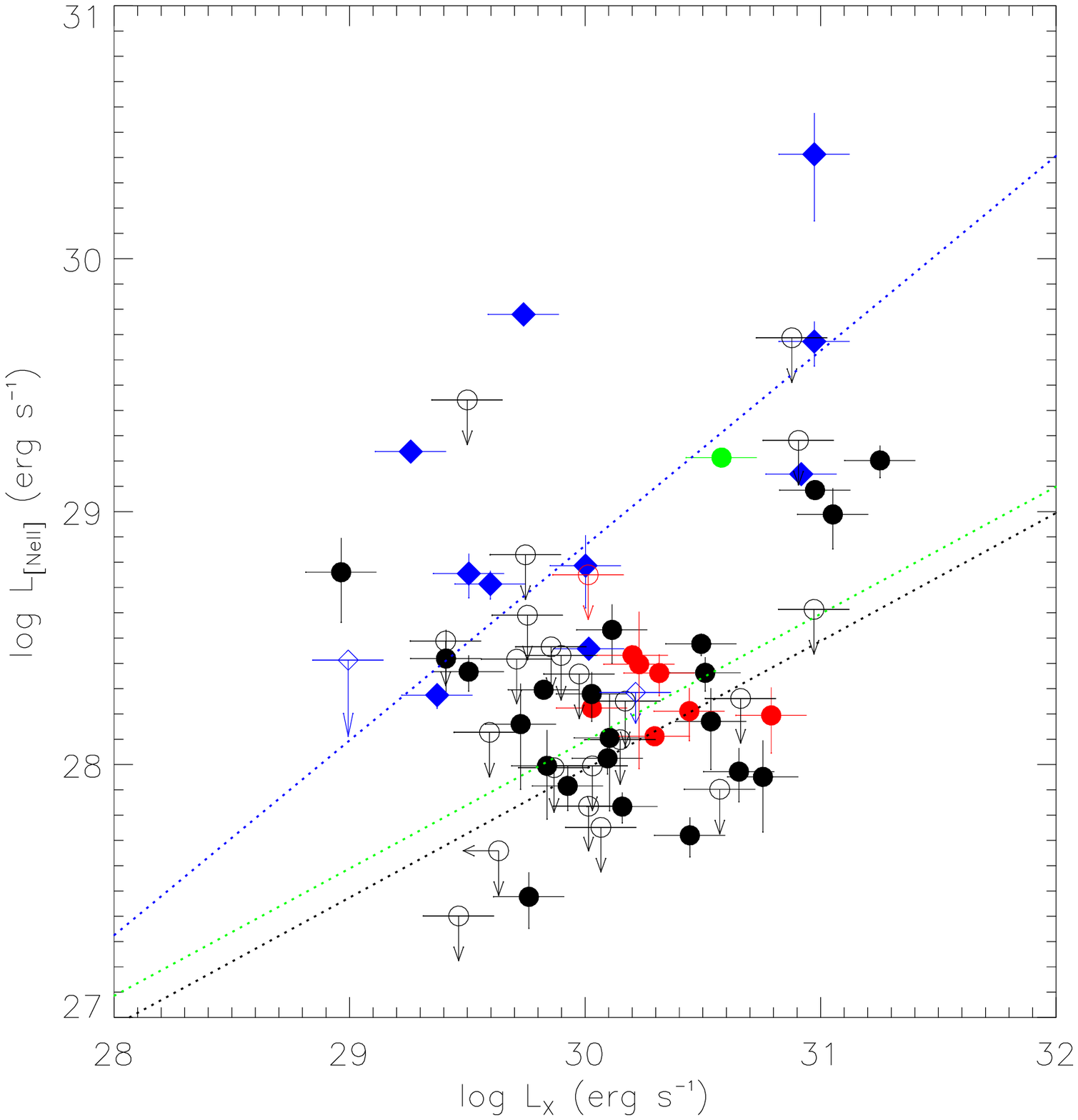}
\caption{$\log L_{\rm [Ne~II]}$ vs $\log L_{\rm X}$. Black circles refer to optically thick disks without jets, red circles to
transition disks, and blue diamonds to jet sources. The green circle marks the position of CS~Cha, a transition disk with
a jet. Filled and open symbols refer to detections and non-detections, 
respectively. Error bars comprise a factor of 2 for X-rays (typical for the range of variability), but represent the 
actual measurement  and fit uncertainties for
$L_{\rm [Ne~II]}$. Regression lines are given for the entire sample (green) and also separately for the non-jet objects (black,
including transition disks) and the jets (blue). Four objects defining the upper envelope of the distribution
have been excluded from the regression analysis of the {\it combined sample}, namely  T~Tau S, DG~Tau, EC~82, and Sz~102.  
         }
\label{fig2} 
\end{figure}

We first discuss a possible dependence between $L_{\rm [Ne\,II]}$ and the intrinsic, unabsorbed $L_{\rm X}$. Fig.~\ref{fig2} relates 
the two quantities for the entire sample, distinguishing between the three object classes (using different
symbol shapes and colors), with separate (open) symbols for upper limits (mostly in $L_{\rm [Ne\,II]}$). We provide
error bars for $L_{\rm [Ne\,II]}$  as derived from spectral analysis, while for X-rays spectral-fit errors are
normally not relevant as the range of uncertainty is dominated by variability on various time scales. Such 
variability is, apart from singular flares, typically characterized by flux variations within a factor of two from 
low to high levels. We therefore adopted error bars defining flux deviations of $\sqrt{2}$ to both higher and lower 
values.

Note the large range now available in both variables, amounting to $\approx$2~dex in $L_{\rm X}$  and $\approx$3~dex in  $L_{\rm [Ne\,II]}$,
i.e., much wider ranges than in previous studies \citep{espaillat07, lahuis07, pascucci07}. 
No sharp correlation is found although a statistically significant dependence exists after excluding the four 
very strong [Ne\,{\sc ii}] detections (for T~Tau S, DG~Tau, Sz~102, and EC~82) that define the upper envelope of 
the distribution (a correlation still exists if they are included). Three of these objects eject prominent 
jets (T~Tau S, DG~Tau, and Sz~102) while EC~82
is a little studied object with a relatively strongly absorbed high-inclination/near-edge-on disk 
\citep{pontoppidan05}. The best-fit regression for the remaining sample has a slope of $0.50\pm 0.15$, with 
a  low probability, $P\la 6\%$, for this correlation being attained by chance (Table~\ref{table6}). 

\begin{table*}[t!]
\caption{Correlation summary}
\begin{tabular}{lrrrrrrr}
\hline
\hline
Parameter & \# points & \# upper limits & $P$(Cox Hazard) & $P$(Kendall tau) & $P$(Spearman rho) & Slope & Intercept \\
\hline
$\log L_{\rm X, 0.3-10, unabs}^a$ & 60 & 24 & {\bf 0.016}  & {\it 0.055}  & {\it 0.059}  & 0.50$\pm$0.15 & 12.99 \\
 \quad only jets$^a$              & 12 &  2 & {\bf 0.015}  & {\bf 0.027}  & {\bf 0.037}  & 0.77$\pm$0.27 & 5.75  \\
 \quad without jets$^a$           & 50 & 22 & {\bf 0.011}  & {\bf 0.043}  & {\it 0.056}  & 0.51$\pm$0.14 & 12.77 \\
$\log L_{\rm X, 0.3-10, abs}^a$   & 60 & 24 & {\bf 0.020}  & {\bf 0.029}  & {\bf 0.050}  & 0.51$\pm$0.17 & 13.02 \\
 \quad only jets$^a$              & 12 &  2 & {\bf 0.031}  & {\bf 0.041}  & {\it 0.076}  & 0.78$\pm$0.26 & 5.84  \\
 \quad without jets $^a$          & 50 & 22 & {\bf 0.022}  & {\it 0.072}  & {\it 0.091}  & 0.44$\pm$0.18 & 15.06 \\
$\log \dot{M}_{\rm acc}$          & 36 &  9 & 0.273        & {\bf 0.037}  & {\bf 0.046}  & 0.35$\pm$0.11 & 31.14 \\
 \quad only jets                  & 12 &  2 & 0.789        & 0.782        & 0.637        & 0.28$\pm$0.27 & 30.90 \\
 \quad without jets               & 24 &  7 & 0.181        & 0.639        & 0.717        &-0.05$\pm$0.07 & 27.70 \\
$\log (\dot{M}_{\rm acc}L_{\rm X})$& 33&  9 & {\it 0.053}  & {\bf 0.012}  & {\bf 0.021}  & 0.44$\pm$0.10 & 18.59 \\
 \quad only jets                  & 11 &  2 & {\it 0.080}  & {\it 0.059}  & {\bf 0.048}  & 0.45$\pm$0.16 & 18.51 \\
 \quad without jets               & 22 &  7 & 0.431        & 0.942        & 0.934        &-0.02$\pm$0.09 & 28.66 \\
EW(H$\alpha$)                     & 55 & 18 & 0.470        & 0.223        & 0.293        & 0.26$\pm$0.18 & 27.82 \\
 \quad only jets                  & 12 &  2 & 0.605        & 0.836        & 0.789        & 0.11$\pm$0.55 & 28.63 \\
 \quad without jets               & 43 & 16 & 0.596        & 0.889        & 0.804        &-0.04$\pm$0.14 & 28.15 \\
EW(O\,{\sc i}$_t$)                & 31 &  9 & {\bf 0.004}  & {\bf 0.008}  & {\bf 0.007}  & 0.66$\pm$0.17 & 28.37 \\
EW(O\,{\sc i}$_t$)$L_{\rm X}$    & 25  &  8 & {\bf 0.002}  & {\bf 0.003}  & {\bf 0.009}  & 0.78$\pm$0.15 &  5.01 \\
EW(O\,{\sc i}$_f$)               & 17  &  4 & {\it 0.091}  & 0.139	  &  0.137	 & 0.35$\pm$0.13 & 28.49 \\ 
EW(O\,{\sc i}$_f$)$L_{\rm X}$    & 16  &  4 & 0.121        & 0.118	  &  0.140	 & 0.45$\pm$0.16 & 15.15 \\
$\log L_{{\rm OI}_t}$             & 28 &  8 & {\bf 0.015}  & {\bf 0.025}  & {\bf 0.034}  & 0.42$\pm$0.10 & 29.72 \\
$\log (L_{{\rm OI}_t}L_{\rm X}$)  & 24 &  7 & {\bf 0.007}  & {\bf 0.015}  & {\bf 0.027}  & 0.46$\pm$0.10 & 16.47 \\
$\log L_{{\rm OI}_f}$             & 17 &  4 & 0.208        & 0.139        & 0.176        & 0.23$\pm$0.09 & 29.41 \\ 
$\log (L_{{\rm OI}_f}L_{\rm X}$)  & 16 &  4 & 0.179        & 0.118        & 0.137        & 0.33$\pm$0.12 & 20.03 \\   
$\log \dot{M}_{\rm loss}$         & 13 &  3 & 0.110        & {\it 0.067}  & {\it 0.100}  & 0.44$\pm$0.14 & 32.06 \\
$\log (\dot{M}_{\rm loss}L_{\rm X}$)& 12& 3 & {\it 0.088}  & 0.215        & 0.263        & 0.46$\pm$0.13 & 18.48 \\
$\log N_{\rm H}$                  & 64 & 24 & 0.206        & 0.157        & 0.192        & 0.28$\pm$0.18 & 28.35 \\
$\log (N_{\rm H}L_{\rm X})^a$     & 60 & 24 & {\it 0.051}  & 0.104        & 0.126        & 0.27$\pm$0.10 & 20.18 \\
\noalign{\smallskip}\hline
\end{tabular}\\
\label{table6}            
Notes: The number of points (``\# points'') includes upper limits, detailed in the column ``\# upper limits''.\\
$L_{\rm X}$ refers to $L_{\rm X, 0.3-10, unabs}$. $P$ is the probability that the correlation is obtained by chance. $P$ values
up to 5\% and between 5\% and 10\%  are printed in boldface and italics, respectively.\\
$^a$ Four high-$L_{\rm [Ne\,II]}$ objects have not been considered for these correlations, namely T~Tau S, DG~Tau, EC~82, and Sz~102
(see text and figures for details). Sz~102 has not been considered in any of the correlations that involve $L_{\rm X}$.
\end{table*}

As indicated above, the transition disks behave like the optically thick disks without jets. The jet sources,
in contrast (shown as blue diamonds in Fig.~\ref{fig2} and further figures), are systematically more luminous in 
[Ne\,{\sc ii}], revealing only modest overlap with the region
occupied by the other objects. A separate regression analysis for the jet sources indicates a significant 
dependence with a regression slope of $0.77\pm 0.27$, i.e., compatible with proportionality.  The dependence
is less tight for the non-jet objects although still significant, with a shallower slope of $0.51\pm 0.14$. 
This trend is shallower than what simple theories would predict, i.e. trends close to proportionality (\citealt{meijerink08,
hollenbach09}, see Sect.~\ref{nedisks}).

\subsection{[Ne\,{\sc ii}] emission and accretion}
\begin{figure}
\hbox{
\includegraphics[angle=0,width=9.cm]{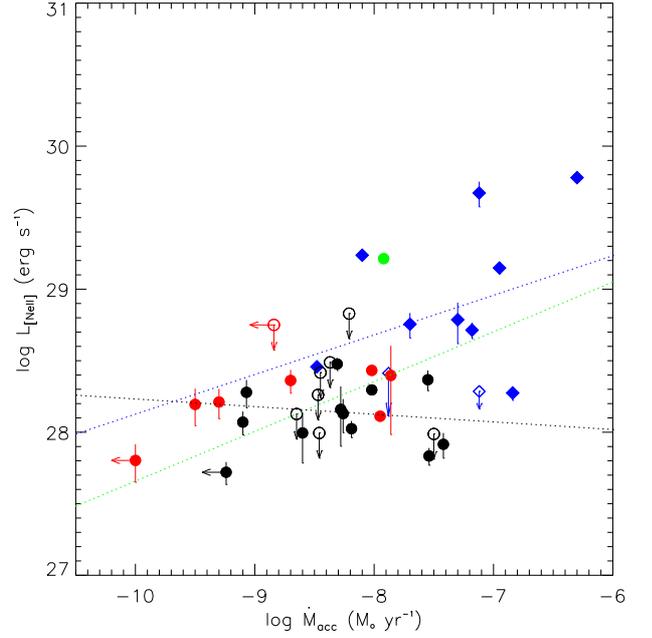}
}
\caption{$\log L_{\rm [Ne~II]}$ vs $\log \dot{M}_{\rm acc}$. Symbols, lines and colors are as in Fig.~\ref{fig2}.
          Regression analysis excludes the upper limits to $\dot{M}_{\rm acc}$ for  CoKu~Tau~4, DoAr~25, and SR~21.
         }
\label{fig3} 
\end{figure}

Fig.~\ref{fig3} relates $L_{\rm [Ne\,II]}$ to the mass accretion rate, as suggested by \citet{espaillat07}. Again,
a large range of $\dot{M}_{\rm acc}$ values is covered, spanning the interval of $\approx 10^{-10}- 10^{-6}~M_{\odot}$~yr$^{-1}$.
No  correlation is evident among the jet sources or the disks without jets separately,  with $P \approx 18-79$\%. 
However, stronger accretors are predominantly jet sources, and they reveal higher $L_{\rm [Ne\,II]}$. 

A dependence between the two variables is therefore ambiguous. Although a  physical dependence may be 
absent, the segregation into objects with and without jets may produce an apparent correlation. 
Jet engines are typically younger and more active objects, and given a rough correlation between
accretion rate and outflow rate (e.g., \citealt{hartigan95}), jet sources typically also show
high accretion rates. Our separate finding that jet sources are generally more luminous [Ne\,{\sc ii}] sources
thus may in fact produce a {\it bi-modal distribution} rather than a correlation based on any physical 
dependence.

We have also correlated  $L_{\rm [Ne~II]}$  with the equivalent width of the H$\alpha$ line (an accretion 
indicator, Table~\ref{table4}), but found no significant trend (Table~\ref{table6}).

\subsection{[Ne\,{\sc ii}] emission and [O\,I] emission}
\begin{figure}
\includegraphics[angle=0,width=9.cm]{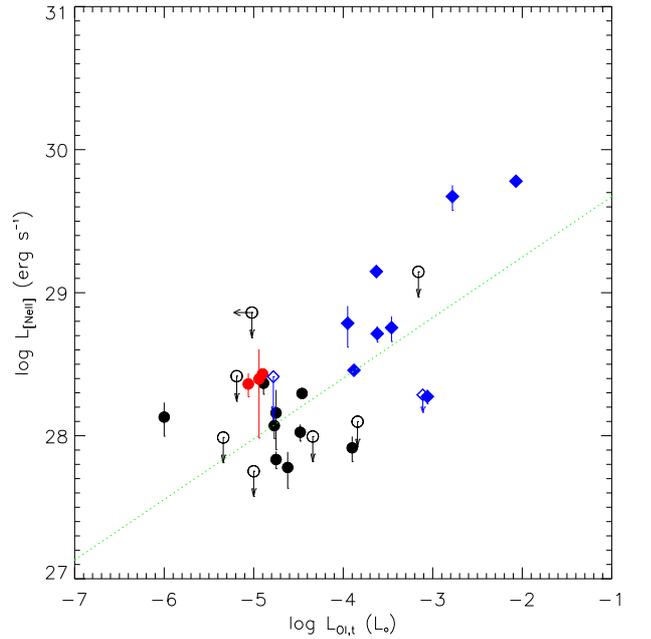}
\caption{$\log L_{\rm [Ne~II]}$ vs $\log L_{{\rm [O~I]},t}$ for the entire [O\,{\sc i}] line flux.  
 The upper limits to $\log L_{{\rm [O~I]},t}$ for HT Lup has been excluded from the regression fit. 
 Symbols, lines and colors are as in Fig.~\ref{fig2}.}
\label{fig4} 
\end{figure}

Motivated by the systematic difference between $L_{\rm [Ne\,II]}$ for jet sources and objects without jets,
we now introduce parameters that diagnose - possibly among other things - mass loss, namely the equivalent width 
and the luminosity of the [O\,{\sc i}]~$\lambda$6300 line. We first study whether [O\,{\sc i}] and 
[Ne\,{\sc ii}] emission correlate in a general way, regardless of their origin, perhaps suggesting that they
are diagnostic lines of {\it the same} emission regions. We indeed find a significant correlation ($P\la 3$\%) 
between the respective luminosities (Fig.~\ref{fig4}, Table~\ref{table6}).

 Although the subsamples in consideration are 
naturally dominated by stars with strong mass loss, i.e., objects with jets, many further CTTS show spectroscopic
evidence for [O\,{\sc i}] emission, perhaps resulting from small, spatially unresolved jets
\citep{hartigan95, hirth97} but also from disk surface layers \citep{hartigan95, acke05, meijerink08}.

\subsection{[Ne\,{\sc ii}] emission and mass loss indicators}
The origin of [O\,{\sc i}]~$\lambda$6300 line emission is ambiguous; the low-velocity
component has been interpreted as being formed at the disk surface, in analogy to [Ne\,{\sc ii}] emission 
\citep{hartigan95, acke05, meijerink08}. For this low-velocity component, thus, a correlation between [Ne\,{\sc ii}] and [O\,{\sc i}] luminosities would directly support the disk emission model
for  [Ne\,{\sc ii}]. The high-velocity component is - less ambiguously - ascribed to a jet \citep{hartigan95, hirth97}. Again,
a correlation would support  a model in which shocks or X-ray irradiation from the star excite the [Ne\,{\sc ii}] line in
the jet gas. The statistics at hand is, unfortunately, insufficient to draw convincing conclusions with respect to correlations 
with the [Ne\,{\sc ii}] luminosity. 

\begin{figure}
\includegraphics[angle=0,width=9.cm]{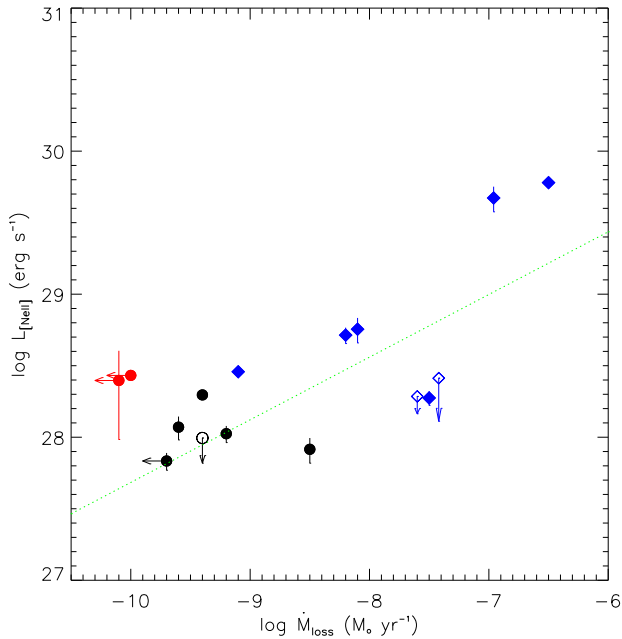}
\caption{$\log L_{\rm [Ne~II]}$ vs $\dot{M}_{\rm loss}$. 
 For the regression analysis, objects with upper limits in $\dot{M}_{\rm loss}$, namely BP Tau, FM Tau, V836 Tau, GM Aur, were not considered.
 Symbols, lines and colors are as in Fig.~\ref{fig2}.}
\label{fig5} 
\end{figure}

We plot in Fig.~\ref{fig5} $L_{\rm [Ne~II]}$ against the mass loss rate, $\dot{M}_{\rm loss}$, as determined from [O\,{\sc i}] luminosities \citep{hartigan95}.
There is a clear trend (at the 10\% probability level, see Table~\ref{table6}), further supporting our view that jets define 
a separate class of [Ne\,{\sc ii}] emitters. The trend suggests that the [Ne\,{\sc ii}] luminosity increases with increasing mass loss
although the relation is non-linear (exponent of $\approx 0.4-0.5$, see Table~\ref{table6}).

\subsection{[Ne\,{\sc ii}] emission and the gaseous environment of CTTS}\label{gasenvironment}
Although we exclude Class I sources from consideration in this work, many CTTS/Class II sources show signs
of excessive absorption (in X-rays) and extinction (in the optical) compared to weak-lined T Tauri stars(WTTS)/Class III objects. For the
XEST survey of the Taurus Molecular Cloud the visual extinction, $A_{\rm V}$, is larger by $\approx 1$ magnitude 
for CTTS compared to WTTS (median; $\approx 2$ magnitudes for the mean), where the median $A_{\rm V}$ for 
WTTS is $\approx 0.9$ magnitudes. There are several possible explanations for this result: i) CTTS are predominantly
located in denser regions of the surrounding molecular cloud; ii) there is more residual gas in the immediate
vicinity of the star.  Additional extinction may arise from  iii) stronger dusty stellar winds and disk winds, 
iv) from the disk 
itself, or  from v) dusty accretion flows. Point iv) is clearly important for near-edge-on disks, with lines-of-sight 
traversing the upper disk layers (e.g., \citealt{kastner05}); such orientation is, however, rare. 
Point (v) is probably less relevant given high temperatures
in the the disk-star accretion flows, resulting in largely dust-depleted gas which is still capable of 
absorbing X-rays (see Sect.~\ref{discussion} below).

Studying [Ne\,{\sc ii}] emission in the context of extinction or X-ray photoelectric absorption may provide important 
hints on its origin. On the one hand, higher levels of gas in the immediate stellar environment absorb
more EUV and X-ray flux, thus suppressing formation of Ne$^{+}$ in the surface layers of the circumstellar disks.
In extreme cases, no EUV or X-ray radiation may reach the disk surface - see Sect.~\ref{discussion}. On the other
hand, [Ne\,{\sc ii}] may be formed in the absorbing layer itself, which would be suggested if $L_{\rm [Ne\,II]}$ increases
with increasing gas columns.

\begin{figure}
\includegraphics[angle=0,width=9.cm]{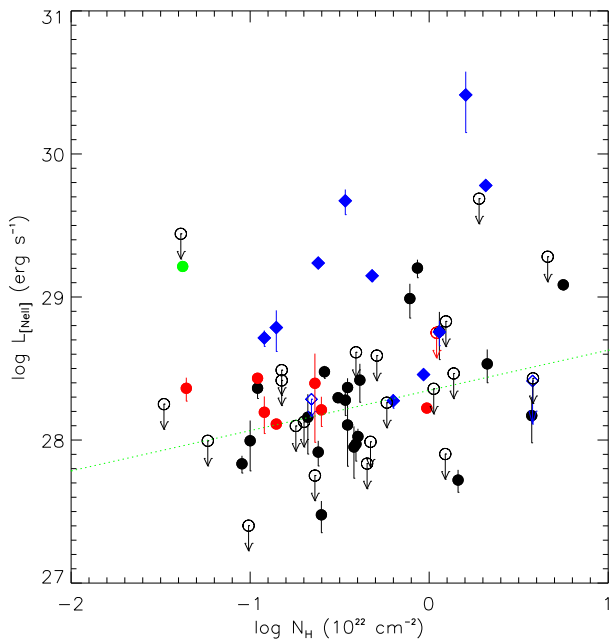}
\caption{$\log L_{\rm [Ne~II]}$ vs  $\log N_{\rm H}$  as determined from X-ray spectral observations along the line of sight to
the star. Symbols, lines and colors are as in Fig.~\ref{fig2}.
 }
\label{fig6} 
\end{figure}

In Fig.~\ref{fig6}, we plot $L_{\rm [Ne\,II]}$  vs the X-ray derived  $N_{\rm H}$
along the line-of-sight to the stellar X-ray source.  
We find a weak but hardly significant trend toward higher $L_{\rm [Ne\,II]}$ with increasing $N_{\rm H}$.

Because the effect of $N_{\rm H}$ is to lower the X-ray flux that reaches a circumstellar disk, we can also
ask whether $L_{\rm [Ne\,II]}$ correlates with the X-ray flux {\it behind the absorbing medium}, i.e., the ``attenuated 
flux''. This flux is difficult
to determine as the absorbing gas column that X-rays or EUV photons encounter {\it toward the disk surface} is unknown.
However, we do know the attenuated flux reaching the observer. Using the
observed, attenuated $L_{\rm X}$ and assuming that the
latter is, on average, in some ways related with the attenuation between the star and the [Ne\,{\sc ii}] emitting
source, the result is very similar to what we derived for the  intrinsic luminosities in Fig.~\ref{fig2}, not permitting
further conclusions.

\subsection{[Ne\,{\sc ii}] emission and the role of $L_{\rm X}$}
\begin{figure}
\includegraphics[angle=0,width=9.0cm]{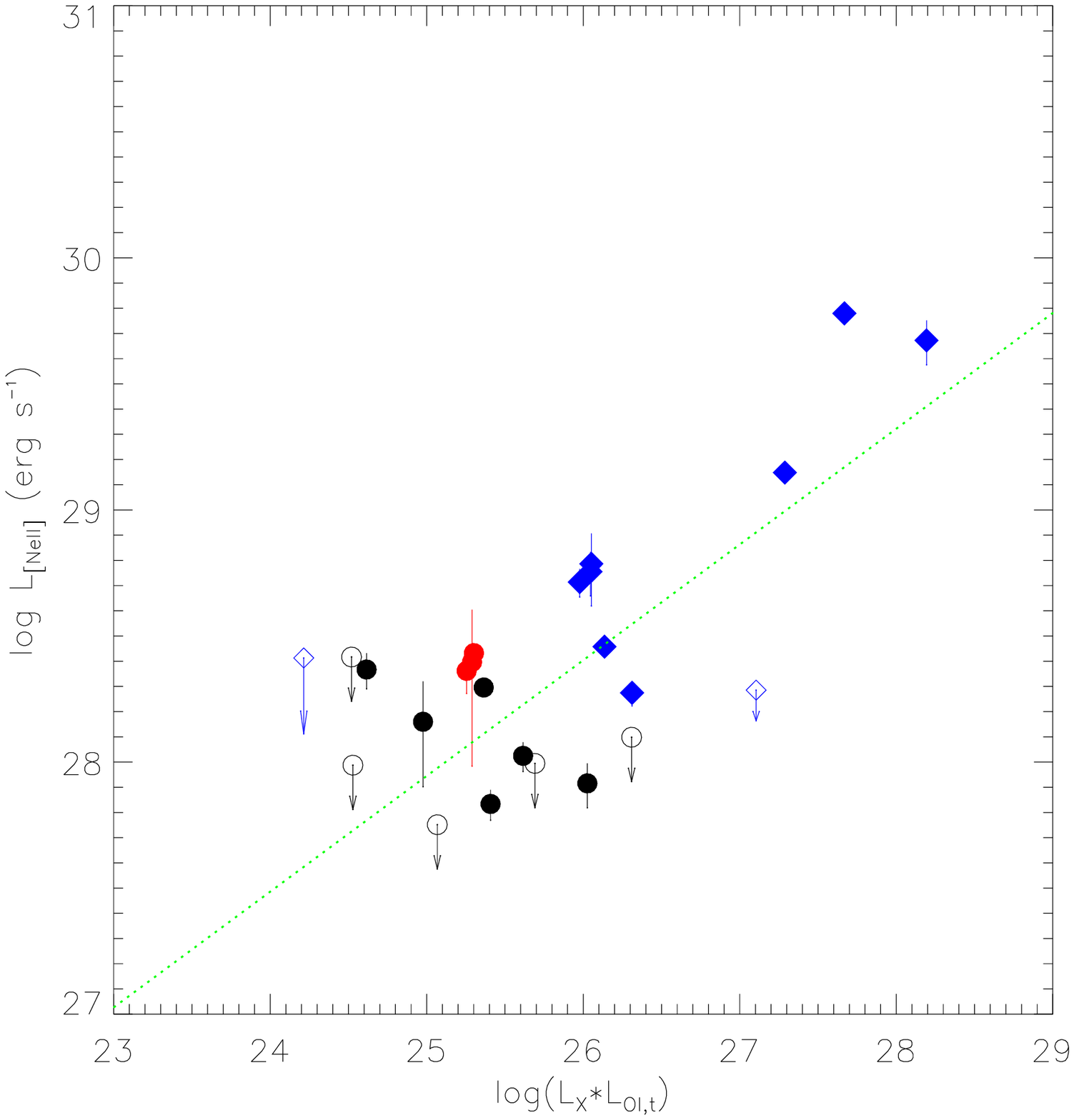}
\includegraphics[angle=0,width=9.0cm]{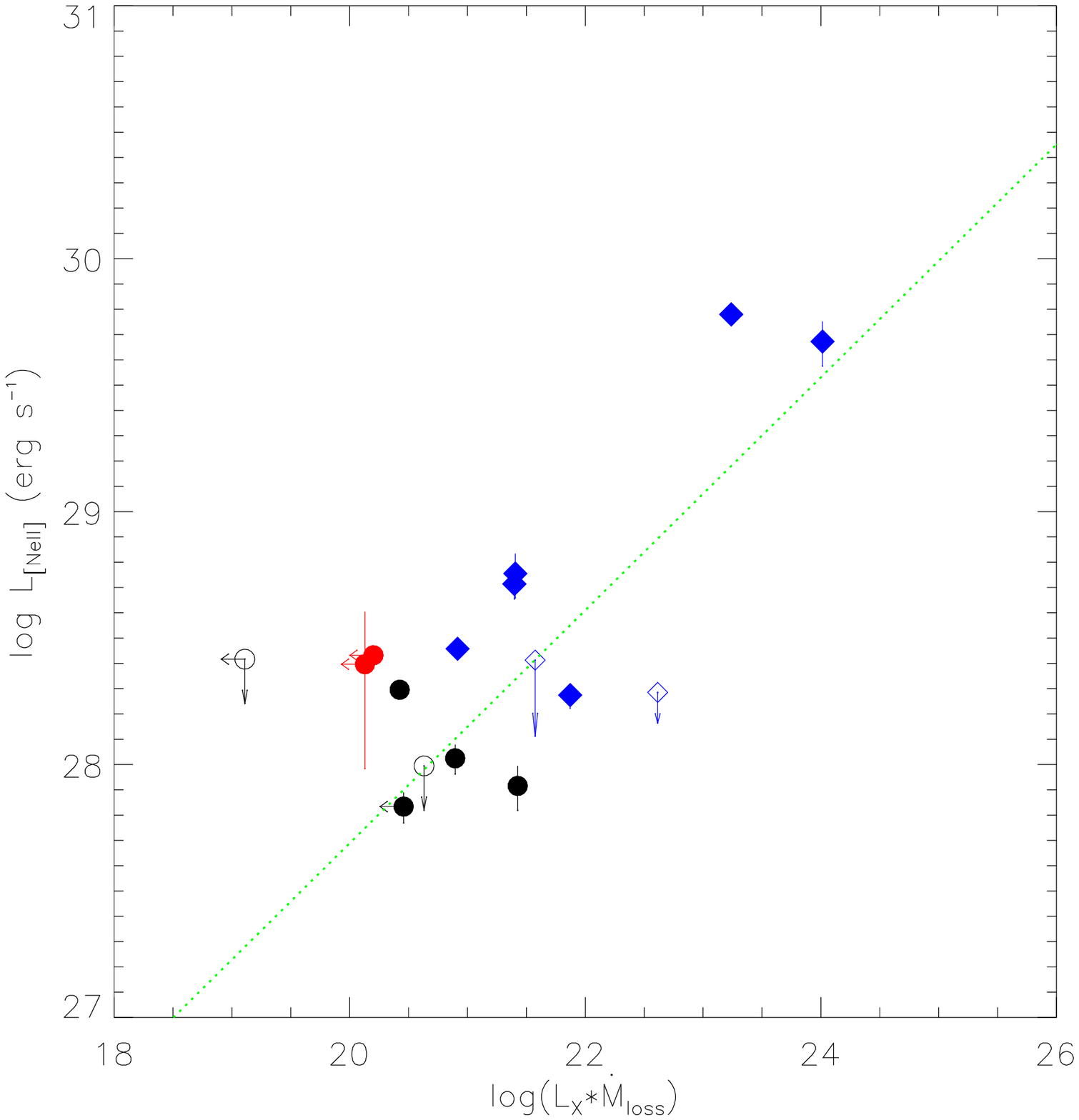}
\caption{$\log L_{\rm [Ne~II]}$ vs the logarithm of the product of  $L_{\rm X}$ times $L_{{\rm [O\,I]},t}$ (upper) and $\dot{M}_{\rm loss}$ (lower). 
	 For the regression analysis, objects with upper limits in $\dot{M}_{\rm loss}$, namely BP Tau, FM Tau, V836 Tau, GM Aur, were not considered.
	 Similarly, the upper limit to $\log L_{{\rm [O~I]},t}$ for HT Lup have been excluded from the regression fit analysis. 
	 Symbols, lines and colors are as in Fig.~\ref{fig2}
 }
\label{fig8} 
\end{figure}

The leading models for [Ne\,{\sc ii}] production posit that the X-ray or EUV flux ionizes and excites Ne. We therefore 
also test whether $L_{\rm [Ne\,II]}$ may depend on some ad hoc products between $L_{\rm X}$ and the parameters related to the [Ne\,{\sc ii}]
emission source. Fig.~\ref{fig8} shows  two examples for the products of $L_{\rm X}$ with 
$L_{{\rm [O\,I],}t}$, and $\dot{M}_{\rm loss}$. The quality of the correlations remains similar (Table~\ref{table6}).

\section{Discussion}\label{discussion}
We suggest (as in \citealt{guedel09a}) four potential [Ne~{\sc ii}] emission regions in stellar environments, partly 
supported by previous work:
\begin{itemize}
\item Disk surface layers irradiated by EUV or X-rays, or heated by accretion shocks; this model has 
      been favored by initial theoretical work \citep{glassgold07};
\item photoevaporative flows above the disk surface, as suggested by \citet{herczeg07} and modeled by \citet{alexander08};
\item jets, as suggested from a statistical sample more thoroughly discussed in the present
      paper, and from explicit observations of  the T Tauri triple  \citep{boekel09}; both X-ray
      induced and shock-induced [Ne\,{\sc ii}] flux production is possible;
\item absorbing accretion columns close to the star.
\end{itemize}

We now discuss our results in the context of these emission models.

\subsection{[Ne\,{\sc ii}] emission from disk surface layers}\label{nedisks}
Heating and ionization of gaseous disk surface layers by stellar X-rays or EUV photons has defined a 
leading theory of [Ne\,{\sc  ii}] production, as initially suggested by \citet{glassgold07} and followed up
by \citet{ercolano08}, \citet{meijerink08}, and \citet{hollenbach09}. Ne$^{+}$ (and Ne$^{++}$) are either 
produced by ejection of a K-shell or an L-shell electron after absorption of an X-ray photon in weakly 
ionized, X-ray heated gas with temperatures of a few 1000~K \citep{glassgold07} or are collisionally ionized in a
hot, strongly ionized ($\approx 10^4$~K) ``H\,{\sc ii}'' region formed at the disk surface after
irradiation by EUV photons \citep{gorti08, hollenbach09}.
 
\citet{glassgold07} estimated  [Ne\,{\sc ii}] fluxes for two disk models in which either mechanical heating
or X-ray heating is dominant, with a stellar $L_{\rm X} = 2\times 10^{30}$~erg~s$^{-1}$, predicting 
typical fluxes of $10^{-14}$~erg~cm$^{-2}$~s$^{-1}$ at a distance of 140~pc, or $\log L_{\rm [Ne\,II]} 
\approx 28.4$. Such luminosities compare very favorably with the bulk of our distribution, although
the most extreme [Ne\,{\sc ii}] luminosities exceed this level by up to two orders of magnitude. 

\begin{figure}
\includegraphics[angle=0,width=9.cm]{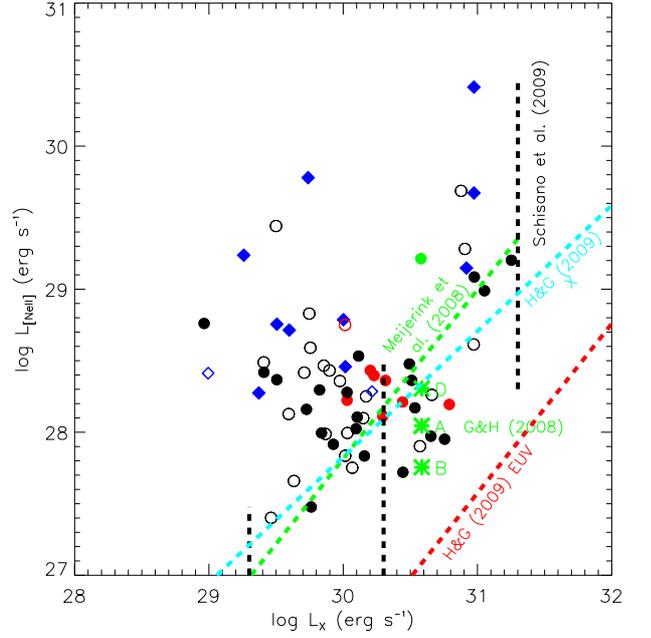}
\caption{Same as Fig~\ref{fig2}, but model predictions for [Ne\,{\sc ii}] disk emission are overplotted, based
on \citet{meijerink08} for a fixed disk (green dashed line), \citet{gorti08} (three purple stars),\citet{schisano09} (vertical dashed black lines), 
and \citet{hollenbach09} (dashed cyan line for X-ray layer, dashed red line for EUV layer).}
\label{fig9} 
\end{figure}

These calculations were extended by \citet{meijerink08} to include various stellar $L_{\rm X}$ values,  $L_{\rm [Ne\,II]}$
being derived from the integrated  emissivities across the entire disk for the case of dominant X-ray heating.
Although the authors mention that their models should not be used to suggest a general correlation 
between the two parameters as the disk properties have been held fixed, these models provide a guide to what
can be expected for similar disks. Their model trend is overplotted in Fig.~\ref{fig9} (dashed green line). It indeed
does provide a good description of the {[Ne\,{\sc ii}] luminosity level} for the optically thick disks,
although the slope of the observed trend is not reproduced, and the jet-driving stars show a much higher $L_{\rm [Ne\,II]}$
than the models.

\citet{ercolano08} estimated $L_{\rm [Ne\,II]}\approx 8.7\times 10^{-7}L_{\odot}$ or  $\log L_{\rm [Ne\,II]} \approx 27.5$
based on radiative transfer calculations assuming $L_{\rm X} = 2\times 10^{30}$~erg~s$^{-1}$. Although this value is
lower than observed at this $L_{\rm X}$ (see Fig.~\ref{fig9}), it matches $\log L_{\rm [Ne\,II]}$ of some of the fainter
objects in our sample. These models were extended to include variations in X-ray spectral hardness  and 
also disk flaring \citep{schisano09}. Disk flaring was modeled by adapting values typically observed in dust disks, 
but also by calculating the hydrostatic, flaring disk structure self-consistently for the
gas component. These calculations demonstrate  that disk flaring is of utmost importance 
(given the largely increasing cross section of a strongly flared disk). Also depending on the spectral hardness 
(based on an {\it absorbed} X-ray spectrum), characteristic values  for $\log L_{\rm [Ne\,II]}$ are 
$2\times 10^{26} - 3\times 10^{27}$~erg~s$^{-1}$ for $L_{\rm X} = 2\times 10^{29}$~erg~s$^{-1}$,
$1\times 10^{27} - 3\times 10^{28}$~erg~s$^{-1}$ for $L_{\rm X} = 2\times 10^{30}$~erg~s$^{-1}$,
$2\times 10^{28} - 3\times 10^{30}$~erg~s$^{-1}$ for $L_{\rm X} = 2\times 10^{31}$~erg~s$^{-1}$ (see Fig.~\ref{fig9}).
Such values again compare favorably with the {\it optically thick} disk sample except for the most [Ne\,{\sc ii}] luminous
objects among them, as already pointed out by \citet{schisano09}.

\citet{gorti08} presented calculations of [Ne\,{\sc ii}] emission from optically thick disks  irradiated by UV, EUV, and X-rays
(with similar luminosities in each band, $\approx 10^{-3}L_{\odot}$; model ``A''). Variants involved a 100 times lower dust
opacity (representing an evolved disk, model ``B''), absence of X-rays (model ``C''), and a tenfold higher FUV luminosity
(model ``D''). The resulting total $L_{\rm [Ne\,II]}$ are plotted for models A, B, and D in Fig.~\ref{fig9}, 
and once again they match the fainter [Ne\,{\sc ii}] fluxes from optically thick disks quite well.

\citet{hollenbach09} estimate
the total [Ne\,{\sc ii}] flux both from the EUV-heated H\,{\sc ii} disk surface layer and the X-ray heated subsurface layer to
find that for plausible stellar EUV and X-ray spectra, the X-ray layer produces $\approx$ twice as much $L_{\rm [Ne\,II]}$
for equal total luminosities in both bands, simply because of the much smaller column available to EUV, given the strong absorption 
by H and He. The numerically calculated $L_{\rm [Ne\,II]}$ values compare well with values from other authors discussed 
above, showing a near-linear increase with $L_{\rm X}$ or $L_{\rm EUV}$, see Fig.~\ref{fig9}, the emission mostly originating
from within 10~AU of the star. The authors also compute [O\,{\sc i}]~$\lambda$6300
luminosities from the EUV and X-ray heated disk layers, concluding, however, that typical luminosities remain orders of magnitude
lower ($\la 10^{-6}L_{\odot}$ to a maximum of $10^{-4}L_{\odot}$) than strong [O\,{\sc i}] luminosities observed in many CTTS 
($10^{-6}L_{\odot} - 10^{-3}L_{\odot}$, referring to the so-called low-velocity component that has been attributed to disk emission; 
\citealt{hartigan95}). 

In summary, most of the model calculations described above yield $L_{\rm [Ne\,II]}$ of the same order
as observed for the {\it optically thick} disk sample, with a tendency for somewhat higher {\it observed} luminosities 
especially  for $L_{\rm X} \la 10^{30}$~erg~s$^{-1}$. We suggest that further features that might enhance  [Ne\,{\sc ii}]
emission are puffed-up disks as well as actively photoevaporating disks, because a larger fraction of the X-rays (or 
EUV radiation) would be absorbed to subsequently produce [Ne\,{\sc ii}] emission. There is indeed evidence
that [Ne\,{\sc ii}] emission may come from a disk-related photoevaporative flow, suggested by small
but non-zero radial velocities measured as a slight blue-shift of the [Ne\,{\sc ii}] line \citep{pascucci09}.

\subsection{Circumstellar absorption: Do X-rays/EUV reach the disk?}\label{absorption}

In the light of the presence of various gaseous components in the immediate stellar vicinity such as accretion columns, disk atmospheres, 
X-winds, or photoevaporative flows, a critical question emerges on whether high-energy photons from the star in fact reach the 
disk surface. This problem has been addressed by \citet{hollenbach09} who present both analytic estimates and
numerical results from a geometrically self-consistent disk model shielded by an outflowing wind. Observationally, the
mass loss rates of such winds scale as $\approx \dot{M}_{\rm acc}$ (e.g., \citealt{hartigan95, white04}) and therefore set limits
to the penetration of stellar FUV, EUV, and X-ray photons to the disk surface. Specifically, the authors estimate that FUV
and 1~keV X-ray photons penetrate the wind for $\dot{M}_{\rm acc} \la 4\times 10^{-7}~M_{\odot}$~yr$^{-1}$, while softer X-rays and 
EUV photons require $\dot{M}_{\rm acc} \la 8\times 10^{-9}~M_{\odot}$~yr$^{-1}$. {\it If} [Ne\,{\sc ii}] emission is induced
by soft X-ray or EUV irradiation of the disk, then obviously winds, and by implication accretion rates, must be sufficiently modest.
Many of our sources, in particular those ejecting jets, violate these conditions, i.e., disk [Ne\,{\sc ii}] emission is unlikely
in these cases, at least if soft X-rays or EUV radiation are responsible for the excitation. 

\begin{figure}
\includegraphics[angle=0,width=9.cm]{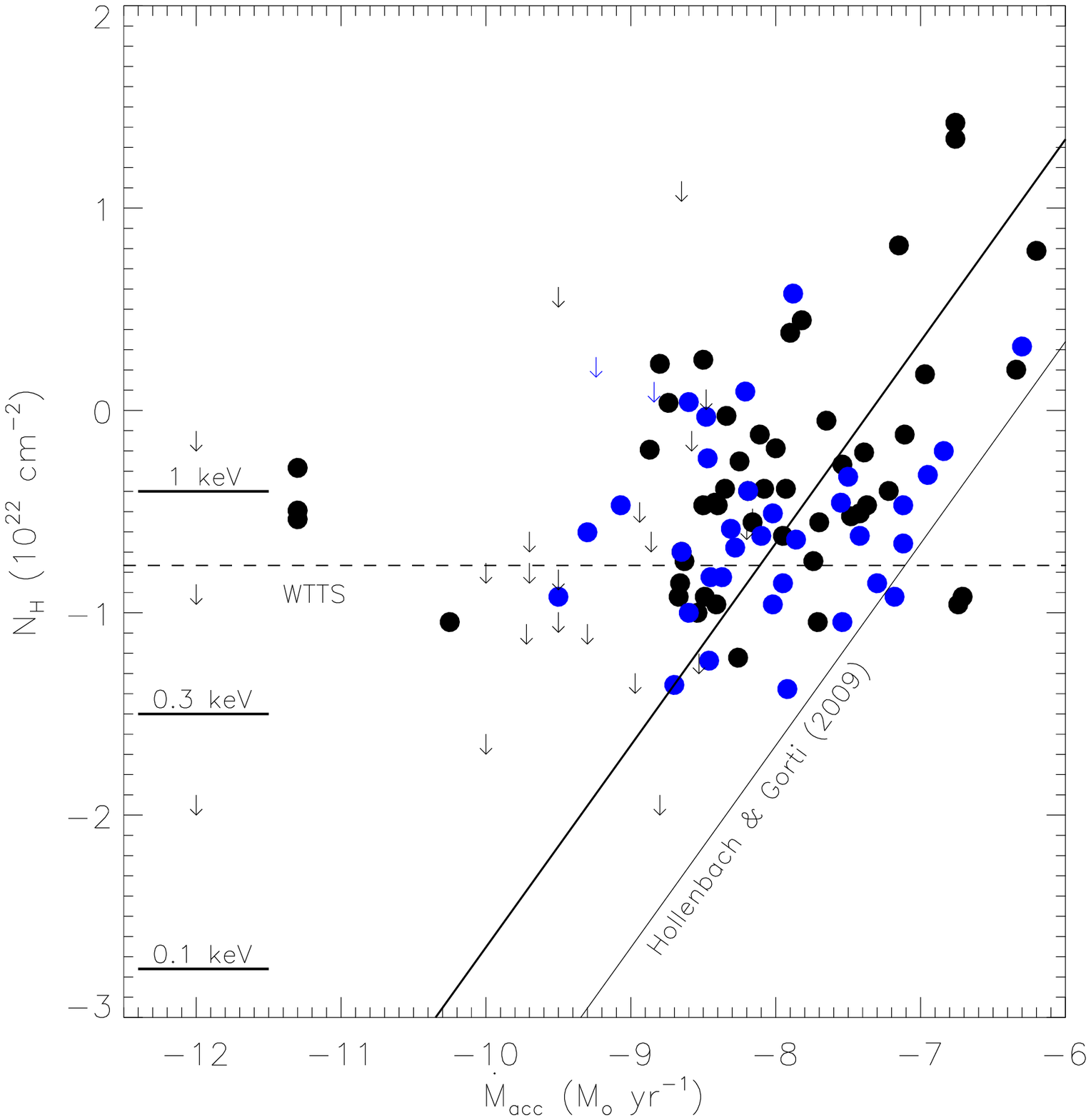}
\caption{$N_{\rm H}$ vs.  $\dot{M}_{\rm acc}$ for CTTS in the present sample (blue symbols) and from the XEST survey
(black symbols; \citealt{guedel07a}). Measurements are shown by bullets, while upper limits to $N_{\rm H}$  are given
by arrows. The thin diagonal line shows a relation adopted by \citet{hollenbach09} for an  accretion-driven wind, while
the  thicker, parallel line shows the same relation shifted upward by a factor of ten.  The dotted horizontal line
shows the (logarithmically) averaged  $N_{\rm H}$  for WTTS in Taurus, and the three short bars on the left indicate
optical depth unity for photons with energies of 0.1~keV, 0.3~keV, and 1~keV.  }
\label{fig9a} 
\end{figure}

What is the observational evidence? As mentioned earlier, the absorbing gas column between the X-ray/EUV emitting corona 
(or accretion spots) and the disk surface is generally unknown but columns are well measured along the line
of sight to the observer. Again assuming that the latter columns on average reflect approximately the values toward the
disk, we plot in Fig.~\ref{fig9a} the X-ray determined $N_{\rm H}$ against the stellar mass accretion rate 
$\dot{M}_{\rm acc}$. Two samples have been used, namely stars from the present work (in blue) and CTTS from the XEST Taurus
X-ray survey \citep{guedel07a} (in black). Upper limits to $N_{\rm H}$  are given by arrows; data for a few objects
for which the spectral fit converged to $N_{\rm H} = 0$ were dropped from consideration. The two samples show
a similar distribution; more importantly, both suggest an increasing trend for $N_{\rm H}$  with increasing $\dot{M}_{\rm acc}$
although - expectedly - with considerable scatter. The trend adopted by \citet{hollenbach09} is shown by a thin line;
it is obviously too low for  $N_{\rm H}$ by about an order of magnitude but otherwise reflects the observed trend well.

Part of the absorption is due to the large-scale gas distribution in the star-forming regions and also interstellar gas along
the line of sight. Because CTTS are systematically more absorbed (or visually extincted) than WTTS \citep{guedel07a}, we
propose that the excess absorption is due to gas in the immediate stellar environment of CTTS, such as 
accretion flows or disk winds. The average of $\log N_{\rm H}$ for WTTS in Taurus is shown by the horizontal line
in Fig.~\ref{fig9a}, corresponding to $N_{\rm H} = 1.7\times 10^{21}$~cm$^{-2}$ (with a standard deviation of the $N_{\rm H}$
distribution of 0.44~dex). Clearly, apart from some scatter, 
almost all CTTS show $N_{\rm H}$  around that value or higher. For $N_{\rm H}$ close to the WTTS level, we cannot determine
its origin but for higher $N_{\rm H}$, we suggest that circumstellar material adds to the absorption.

Fig.~\ref{fig9a} also shows the critical values of $N_{\rm H}$ for which X-rays are attenuated by a factor of $e$, i.e.,
the optical depth unity, for the three photon energies of 0.1~keV, 0.3~keV, and 1~keV (shown by horizontal
bars). For CTTS with excess $N_{\rm H}$, it is clear that EUV radiation is very strongly attenuated, and even soft X-rays
of a few times 0.1~keV cannot reach the disk. For most of these stars, even 1~keV photons are severely attenuated. 

We thus conclude that the most likely high-energy photons reaching the disk and potentially exciting the [Ne\,{\sc ii}] line
are X-ray photons around 1~keV  or higher, and that EUV photons are unlikely to reach any part of the disk, at least
for the sample of CTTS that show  excess $N_{\rm H}$ compared to WTTS. This conclusion would be invalid if a photoevaporative flow
outside the [Ne\,{\sc ii}] producing region is responsible for the absorption; however, calculations by \citet{ercolano09}
show that X-ray driven photoevaporative winds are concentrated in the inner disk, with surface mass-loss rates peaking within
10~AU where [Ne\,{\sc ii}] emission should be most efficient \citep{hollenbach09}. Similarly, X-winds and accretion flows
would absorb photons at radii within the [Ne\,{\sc ii}] emitting disk region.

Of course, excess absorption in CTTS  may turn the absorbing gas itself into an efficient source of [Ne\,{\sc ii}] 
emission as long as the density remains sufficiently low. Testing this hypothesis is, in principle, simple. 
If the absorbing gas itself were a major source of [Ne\,{\sc ii}] emission, then $L_{\rm [Ne\,II]}$ should increase with 
$N_{\rm H}$ or also with $L_{\rm X}N_{\rm H}$ (with the caveat that excessive $N_{\rm H}$ may in some cases be due 
to the disk itself). Although such a trend is present, it is not significant (Table~\ref{table6}).

If X-ray and EUV absorption is important but [Ne\,{\sc ii}] is still predominantly produced by disk irradiation, then 
we would expect  that $L_{\rm [Ne\,II]}$ correlates more closely with the attenuated $L_{\rm X}$ than with the intrinsic 
$L_{\rm X}$, although we recall the caveat that the {\it observed} absorption along our line of sight may differ from the 
absorption between the star and the disk. Furthermore, [Ne\,{\sc ii}] production should {\it decrease} with
increasing $N_{\rm H}$. These two effects have  not been significantly measured (Table~\ref{table6}).

In the most extreme cases, accretion flows shield the circumstellar environment from X-ray or EUV irradiation. This is particularly
evident in strongly accreting sources with X-ray jets as discussed above. In these cases, it is unlikely that X-ray/EUV photons
reach the disk in significant numbers. The origin of the very strong [Ne\,{\sc ii}] emission in these sources remains ambiguous because they all
eject jets that can be very strong [Ne\,{\sc ii}] emitters \citep{boekel09}, but massive, low-density accretion flows attenuating the
X-ray emission may contribute to [Ne\,{\sc ii}] emission as well.

\subsection{[Ne\,{\sc ii}] emission from jets}
One of the principal results of the present study is the evident [Ne\,{\sc ii}] excess in stars with jets, as previously 
suggested by \citet{guedel09a} and exemplified by the study of the T Tau triplet by \citet{boekel09}.
We have been careful not to confuse our sample with strong jets from Class I sources in which other mechanisms
(excitation of [Ne\,{\sc ii}] in the envelope, stronger accretion shocks from material falling onto the disk, etc)
may dominate. On the other hand, jets may also contribute to the generally high level of [Ne\,{\sc ii}] emission
seen in Class I sources reported by \citet{flaccomio09}.
Most of our objects are ordinary CTTS although some extreme cases, such as the flat-spectrum source
DG Tau or the strongly absorbed (by a near-edge-on thick disk) T Tau S have been included. 
Fig.~\ref{fig6} shows that the distribution of
$N_{\rm H}$ is very similar for the jet sources and the optically thick disks without jets (the same is also true for the
transition disks).

What [Ne\,{\sc ii}] emission can be expected from jets? We briefly discuss three principal [Ne\,{\sc ii}] formation
mechanisms, namely from jet shocks, from irradiation of jets by stellar X-rays or EUV flux, and from X-ray emission 
produced by the jets themselves.

\begin{figure}
\includegraphics[angle=0,width=9.cm]{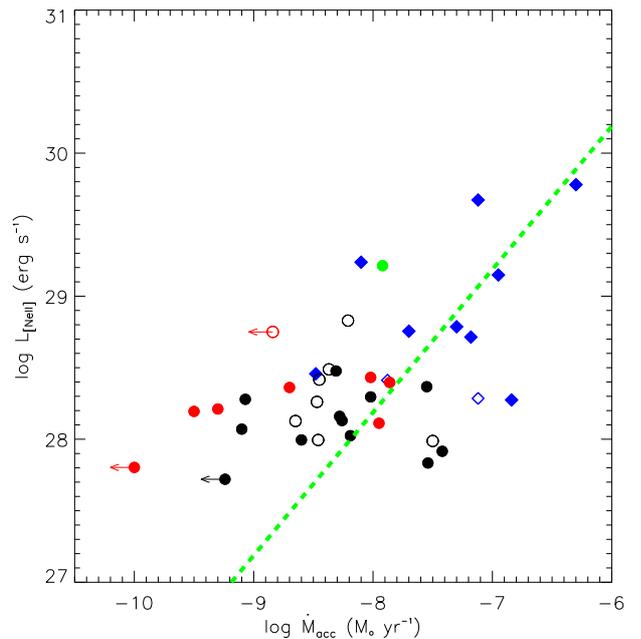}
\caption{Same as Fig~\ref{fig3}, but model prediction for [Ne\,{\sc ii}] jet shock emission is overplotted (dashed green line), referring
to calculations by \citet{hollenbach09} for $\dot{M}_{\rm loss} = 0.1\dot{M}_{\rm acc}$ and all jet gas passing through a shock with
a shock velocity $>100$~km~s$^{-1}$ and pre-shock density $< 10^4$~cm$^{-3}$.  }
\label{fig10} 
\end{figure}

Jet gas is typically heated to $10^4$~K by shocks, shock velocities being a few tens of km~s$^{-1}$ \citep{lavalley00}.
\citet{hollenbach09} estimate $L_{\rm [Ne\,II]}$ from fast ($\ga 100$~km~s$^{-1}$) shocks based
on results from \citet{hollenbach89}, to find a linear relation between $L_{\rm [Ne\,II]}$  and $\dot{M}_{\rm acc}$ (the latter
assumed to scale linearly with $\dot{M}_{\rm loss}$; see Fig.~\ref{fig10}). For typical wind/jet velocities, densities, and mass loss
rates, they find much higher $L_{\rm [Ne\,II]}$ than expected from irradiated disks, fully compatible with our findings
of very high $L_{\rm [Ne\,II]}$ in almost all jet sources.

Jets may also be irradiated by stellar X-rays similar to disk surfaces, and may thus be ionized and heated to
produce  [Ne\,{\sc ii}] emission. Jet irradiation by stellar X-rays is relatively straightforward because  
jets move through wide polar cavities evacuated of much of the circumstellar gas \citep{momose96}.
Evidence for very low gas column densities around CTTS jets has been found from jets that are themselves X-ray
sources (see below) and show very low X-ray attenuation despite the presence of appreciable 
amounts of circumstellar material in other directions \citep{guedel07b}. For the T Tauri system, a simple estimate of
X-ray induced [Ne\,{\sc ii}] line excitation across a lightly  absorbing gas column showed that the observed stellar $L_{\rm X}$ may
indeed yield the observed [Ne\,{\sc ii}] fluxes out to a few arcsec from the star \citep{guedel09a}.

A rather unexpected finding are CTTS jets that produce X-rays themselves \citep{guedel07b, guedel08}. Apart from direct X-ray
imaging of jets (in particular the case of DG~Tau), such jets have also been identified spectroscopically in X-rays. Their anomalous
spectra show a highly absorbed, hard and variable coronal component together with a soft, very weakly absorbed and non-variable
component apparently produced by the jets close to the star. The excessive absorption of the coronal component is an order
of magnitude larger than expected from the  visual extinction of the stellar light if standard gas-to-dust mass ratios are assumed. 
This has been interpreted as being due to dust-depleted accretion streams falling from the disk to the star, thus absorbing 
X-rays from the underlying corona \citep{guedel07b}. It is possible that in these cases the soft jet component is discernible simply
because the stellar component is absorbed at low X-ray energies, while in less strongly accreting (and therefore less absorbed) 
objects the jet component is outshone by the coronal spectrum. Four objects in our sample have been interpreted as showing soft
X-ray jets: DG Tau, DP Tau, HN Tau, and also Sz~102, the latter revealing only a soft component, the hard component possibly
being completely absorbed by a near-edge-on disk \citep{guedel09b}. 

Three of these objects show very high $L_{\rm [Ne\,II]}$, while for DP Tau an upper limit is available. We find no specific
trend for the four objects tighter than what is shown in Figs.~\ref{fig2}, \ref{fig3}, or \ref{fig4}. However, except for Sz~102
where only a soft component is present, we have adopted the hard component as representing the stellar radiation.
The luminosities in the {\it soft} components are, $9.6\times 10^{28}$~erg~s$^{-1}$, $1.5\times 10^{29}$~erg~s$^{-1}$,
$4.0\times 10^{27}$~erg~s$^{-1}$, and $8.9\times 10^{28}$~erg~s$^{-1}$ for DG~Tau, HN~Tau, DP~Tau, and Sz~102, respectively 
(see \citealt{guedel09b}, and this paper for Sz~102). The corresponding $L_{\rm [Ne\,II]}$ values are, respectively,
$6.1\times 10^{29}$~erg~s$^{-1}$,  $5.6\times 10^{28}$~erg~s$^{-1}$, $<2.6\times 10^{28}$~erg~s$^{-1}$, and $1.7\times 10^{29}$~erg~s$^{-1}$,
not suggesting any correlation. However, it may be interesting to note that 
$\dot{M}_{\rm loss}L_{\rm X, soft} = (30, 1.2, <0.15)\times 10^{21}~M_{\odot}$~yr$^{-1}$~erg~s$^{-1}$ for 
DG~Tau, HN Tau, and DP~Tau, respectively, which roughly correlates with $L_{{\rm [O\,I],}f} = (24, 0.6, 0.06)\times 10^{30}$~erg~s$^{-1}$
and with $L_{\rm [Ne\,II]} = (61, 5.6, <2.6)\times 10^{28}$~erg~s$^{-1}$ although the statistics are too small for significant 
conclusions. Although jets may produce both [Ne\,{\sc ii}] emission 
and very soft X-rays independently by shock heating, the latter may also contribute to ionization and heating
of the predominantly cool jet gas locally, thus adding to [Ne\,{\sc ii}] emission. 

Our finding that [Ne\,{\sc ii}] emission is enhanced in CTTS with jets, supported by spatially resolved [Ne\,{\sc ii}] emission
from the T Tau jet system \citep{boekel09}, finds a parallel in observations of infrared rovibrational H$_2$ emission from
similar targets. The H$_2$ $v = 1-0~S(1)$ line at 2.12~$\mu$m shares excitation conditions with [Ne\,{\sc ii}], i.e., 
excitation in warm gas heated by UV, X-rays, or shocks, where emission from the disk gas is expected to be confined within 
30--50~AU (\citealt{beck08} and references therein).  H$_2$ rovibrational emission has been detected from many CTTS, but again,
the emission source is often resolved.  In the \citet{beck08}  high-resolution study of six CTTS
(including DG Tau, T Tau, and RW Aur from our sample), the H$_2$ emission morphologies, its detection beyond 50~AU 
from the star,  excitation temperatures exceeding 1800 K, kinematics measureed in the features, and the consistency with
calculated shock models suggest that the bulk of the H$_2$ emission is  shock-excited emission from jets and outflows
rather than emission from disk gas excited by short-wavelength flux from the central star. A comparison of their H$_2$ map 
of the T Tau system with the spatial distribution of [Ne\,{\sc ii}] emission reported by \citet{boekel09} indeed 
suggests some common emission sources.

On the other hand, our finding of a correlation between [Ne\,{\sc ii}] luminosity and stellar X-ray luminosity
specifically for objects with jets suggests an important role of the stellar short-wavelength radiation in exciting
[Ne\,{\sc ii}] in the jet gas, at least relatively close to the star (see also estimates in \citealt{boekel09} for the
jet system in T Tau detected in [Ne\,{\sc ii}] out to about 2 arcsec). Explicit theoretical calculations by \citet{shang10}
for the X-wind model of a YSO jet irradiated by X-rays supports this conclusion further.

\subsection{Caveats}
Our correlations show  systematic scatter of typically an order of magnitude in $L_{\rm [Ne\,II]}$, regardless
of the parameter against which the latter is plotted. Although some stellar or disk parameters, such as $L_{\rm X}$ or
$\dot{M}_{\rm acc}$ are themselves subject to considerable measurement error, the scatter in $L_{\rm [Ne\,II]}$ clearly
requires further systematic effects. One possibility is that several parameters considered here matter in
concert. We have specifically investigated the correlations with the product of some parameters with $L_{\rm X}$ in
an attempt to show that X-ray irradiation is one important factor to produce [Ne\,{\sc ii}] emission. No decisive
improvement of the correlations was found, however.

On the other hand, we have ignored a number of parameters that may influence [Ne\,{\sc ii}] emission. In particular, we
have not considered disk flaring which can increase the cross section area for stellar X-rays considerably. Numerical
simulations indeed show order-of-magnitude variations as a result of varying disk flaring for otherwise constant
stellar and disk parameters \citep{schisano09}. Another factor is the spectral energy distribution of the ionizing
and heating X-ray source. Again, accurate measurement of the irradiating spectrum is not possible although the intrinsic
stellar X-ray spectrum can be reconstructed from observations. X-ray spectral hardness indeed significantly influences
[Ne\,{\sc ii}] emission from disks in numerical studies \citep{schisano09}. Further factors that have not been considered
in this study include disk gaps and holes (although we showed that our  sample of transition disks behaves similar to
optically thick disks without jets), grain structure and size distribution, the degree of dust settling, or accretion
from the environment onto the disk. Most significantly, unrecognized jets may contribute to some enhancement of  [Ne\,{\sc ii}]
emission, as is clearly evident from the subsample with known jets. Most of these additional features require
further observational study, and some may remain inaccessible.

\section{Conclusions}
We have studied [Ne\,{\sc ii}] emission from a large sample of disk-surrounded pre-main sequence stars, most of them showing optically thick
disks (i.e., Class II sources), but some of them additionally ejecting jets, and others showing signatures of transition disks.
We have been interested primarily in locating the [Ne\,{\sc ii}] emission source, but also in finding clues about the ionization and
excitation mechanism itself. To this end, we have studied correlations between the observed [Ne\,{\sc ii}] line luminosity and stellar and
circumstellar properties such as the stellar X-ray luminosity, the accretion rate, the mass loss rate, and the luminosity in the 
[O\,{\sc i}]~$\lambda$6300 line. Our principal findings can be summarized as follows:

\begin{itemize}

\item Stars ejecting jets are systematically more luminous in  [Ne\,{\sc ii}], statistically by about one order of magnitude. The difference
between jet sources and non-jet sources is in fact so large that the distribution is nearly bi-modal in $L_{\rm [Ne\,II]}$, with 
only small overlap between the jet and non-jet sample. The single case of a transition disk with a jet included in this sample (CS~Cha)
clearly follows the trends for the jet sources, which is expected if the jets produce typically tenfold higher [Ne\,{\sc ii}] emission
than disks. 

\item A weak correlation is present between $L_{\rm [Ne\,II]}$ and $L_{\rm X}$ in the sense that  the strongest X-ray
sources tend to be strong [Ne\,{\sc ii}] sources, and the lowest-luminosity [Ne\,{\sc ii}] sources tend to be X-ray weak,
but systematic scatter probably due to unrelated properties in the sources dominates. The correlation is strongest
for the jet-driving sources.

\item A weak correlation is also present between $L_{\rm [Ne\,II]}$ and the mass accretion rate, $\dot{M}_{\rm acc}$. However,
such a correlation is not recovered separately for the jet sources and the stars without jets. The best-fit trend  between 
$L_{\rm [Ne\,II]}$  and $\dot{M}_{\rm acc}$ is flat for stars without jets.
On the other hand, because the jet sources (with their implied high mass loss rates) tend to show both larger accretion rates and 
higher $L_{\rm [Ne\,II]}$, an apparent correlation is found that is best explained by the bi-modality of the distributions.
We therefore prefer an explanation in which the {\it mass outflow rate} is the relevant parameter.

\item Correlating $L_{\rm [Ne\,II]}$ with outflow and jet indicators such as EW([O\,{\sc i}]), $L_{\rm [O\, I]}$, or $\dot{M}_{\rm loss}$,
we find several significant correlations, in particular when considering that estimates of $\dot{M}_{\rm loss}$ are subject to large
uncertainties, and equivalent widths relate to the underlying stellar continuum. Note that [O\,{\sc i}] emission may also originate in
disk surfaces, but trends are also seen separately for the jet-related, high-velocity component of the [O\,{\sc i}] line.
The trends coherently include jet/outflow sources across a wide range of mass loss parameters and [Ne\,{\sc ii}] emission.

\item Previous theoretical and numerical estimates of [Ne\,{\sc ii}] fluxes from X-ray/EUV irradiated {\it disk} surfaces agree with typical
[Ne\,{\sc ii}] fluxes from stars without jets, but they cannot explain the strong [Ne\,{\sc ii}]  emission from a subsample of putative non-jet 
objects, and they fail explaining almost all [Ne\,{\sc ii}]  sources from stars with jets. The latter are one to two orders 
of magnitude more luminous in  [Ne\,{\sc ii}] than predicted by disk models. As shown by \citet{hollenbach09}, shocks in jets easily
predict such fluxes. It is well possible that several of the stronger [Ne\,{\sc ii}] sources also eject hitherto unrecognized jets.

\item Considering the absorbing gas column density excesses for CTTS (with respect to WTTS), we suggest that additional gas
      in accretion streams or disk winds is responsible. This gas will attenuate soft X-rays and EUV photons sufficiently to
      prevent them from reaching the disk surfaces. If [Ne\,{\sc ii}] is produced in disks, then the exciting radiation is 
      predominantly X-rays with energies of order 1~keV.

\item Transition disks behave, in many ways, like normal optically thick disks without jets. Their [Ne\,{\sc ii}] luminosity is 
bounded by $3\times 10^{28}$~erg~s$^{-1}$. The similar behavior to optically thick disks probably results from
the easy excitation of the [Ne\,{\sc ii}] transition even in small amounts of gas \citep{glassgold07} at column densities
$< 10^{21}$~cm$^{-2}$ \citep{meijerink08}; the total amount of gas in a disk is therefore of little relevance. Because 
the transition is excited out to distances of about 25~AU \citep{meijerink08}, inner holes in transition disks may also not 
be of much relevance; on the contrary, reduced winds from the thin inner disk \citep{hollenbach09} or less disk gas mass in the 
way toward larger distances may alleviate excitation of the [Ne\,{\sc ii}] transition in more extended gas layers.

\item The bi-modality of our distributions, with jet sources being substantially more luminous in [Ne\,{\sc ii}] than non-jet sources,
       suggests that two different emission mechanisms contribute to [Ne\,{\sc ii}] emission. There seems to be little doubt that
       the presence of jets favors strong [Ne\,{\sc ii}] detections, and the most  likely process is [Ne\,{\sc ii}] emission from
       the jets themselves, excited either by shocks or also the stellar X-rays at least in the vicinity of the star \citep{shang10}. 
       The latter mechanism
       is supported by a correlation between $L_{\rm [Ne\,II]}$ and $L_{\rm X}$ for jet sources (Fig.~\ref{fig2}). On the other hand,
       non-jet sources fail to follow some trends seen for jet sources; most evidently, $L_{\rm [Ne\,II]}$ does not 
       correlate with $\dot{M}_{\rm acc}$ for these objects, but rather shows a flat distribution in this parameter (Fig.~\ref{fig3}, see
       also Fig.~\ref{fig10}).        

\end{itemize}
We conclude that jet ejection leads to enhanced [Ne\,{\sc ii}] emission, although we cannot demonstrate whether [Ne\,{\sc ii}] emission
forms in shocks close to the star, or in absorbing accretion flows toward the star as the increased mass loss rates of jet engines
also implies  higher accretion rates, and therefore more massive accretion flows close to the stellar X-ray source.

We still do find indications that the production of [Ne\,{\sc ii}] emission weakly scales with the X-ray luminosity. This finding supports
previous theoretical models of X-ray irradiated and ionized stellar environments although irradiated winds, accretion flows and jets
should also be considered as targets, apart from the so far favored disk surface layers. The correlation found between [Ne\,{\sc ii}]
emission from jet-driving CTTS and $L_{\rm X}$ suggests an important role of stellar short-wavelength radiation in exciting
this line in the jet/outflow gas at least relatively close to the stars themselves.

There is obvious need for deeper studies to disentangle the various possible origins of [Ne\,{\sc ii}] emission. The 
{\it Spitzer} beam is large and potentially includes both unresolved outflow and disk contributions.  Narrow slit 
observation using high resolving power, possibly stepped across the source for integral field spectroscopy, could 
uncover disk line profiles (symmetric, centered at stellar velocity, disk-like velocity range if bound or blue-shifted 
if photoevaporating)  separately from outflow signatures (asymmetric lines, blue-shifted or red-shifted, with high 
velocities) at larger distances from the star. Such observations have now provided first interesting results 
\citep{boekel09, najita09, pascucci09}.

\begin{acknowledgements}
We thank an anomymous referee for helpful comments that improved our paper.
MG thanks the Max-Planck-Institute for Astronomy (Heidelberg, Germany), and Leiden Observatory/Leiden University (Leiden, NL), for 
support during his Sabbatical visit when the study presented here was started.
This work is based [in part] on observations made with the Spitzer Space Telescope, which is operated by the Jet Propulsion 
Laboratory, California Institute of Technology under a contract with NASA.
It is also partly  based on observations obtained with {\it XMM-Newton}, an ESA science mission
with instruments and contributions directly funded by ESA member states and the USA (NASA).
The {\it Chandra X-Ray Observatory Center}  is operated by the Smithsonian Astrophysical Observatory  for and on behalf of
NASA under contract NAS8-03060. Some X-ray data used for the present work were obtained in the framework of projects
supported by NASA grant NNX07AU30G.
This research has made use of the SIMBAD database, operated at CDS, Strasbourg, France.
This publication makes use of data products from the Two Micron All Sky Survey, which is a joint project of the University 
of Massachusetts and the Infrared Processing and Analysis Center/California Institute of Technology, funded by NASA 
and the US National Science Foundation. We have also made use of the ASURV statistical software package maintained by Penn State.
\end{acknowledgements}

\Online

\setcounter{table}{0}
\begin{table*}
\caption{Targets}
\begin{tabular}{llrrr}
\hline
\hline
Star                    & Alternative                                                                                     &RA(J2000.0)& dec(J2000.0)              & dist.	\\					 
                        & names                                                                    			  & (h m s)  & (deg $\arcmin$ $\arcsec$)  & (pc)        \\
\hline
RNO~15                  & HBC~339, CoKu~NGC~1333/2                                                                        & 03 27 47.7  &  30 12 04.3    &   250  \\				      
LkH$\alpha$~270         & HBC~12, NGC~1333~IRS~2                                                   			  & 03 29 17.7  &  31 22 45.1 	 &   250  \\
LkH$\alpha$~271         & HBC~13                                                                   			  & 03 29 21.9  &  31 15 36.4 	 &   250  \\
LkH$\alpha$~326         & HBC~14                                                                   			  & 03 30 44.0  &  30 32 46.7 	 &   250  \\
LkH$\alpha$~327         & HBC~15, IRAS~03304+3100                                                  			  & 03 33 30.4  &  31 10 50.5 	 &   250  \\
LkH$\alpha$~330         & HBC~20, IRAS~F03426+3214                                                 			  & 03 45 48.3  &  32 24 11.9 	 &   250  \\
IRAS 03446+3254         & IRS~4                                                                    			  & 03 47 47.1  &  33 04 03.4 	 &   250  \\
BP~Tau                  & HBC~32, HD~281934, IRAS~04161+2859                                       			  & 04 19 15.8  &  29 06 26.9 	 &   140  \\
FM~Tau                  & HBC~23, Haro~6-1                                                                                & 04 14 13.6  &  28 12 49.2    &   140  \\									      
T~Tau N                 & HBC~35, HD 284419, IRAS~04190+1924                                       			  & 04 21 59.4  &  19 32 06.4 	 &   140  \\
T~Tau S                 & IRAS~04190+1924                                                          			  & 04 21 59.4  &  19 32 06.4 	 &   140  \\
LkCa~8                  & HBC~385, IP~Tau                                                          			  & 04 24 57.1  &  27 11 56.4 	 &   140  \\
DG~Tau                  & HBC~37, Ced~33, IRAS 04240+2559                                          			  & 04 27 04.7  &  26 06 16.3 	 &   140  \\
IQ~Tau                  & HBC~41, LkH$\alpha$~265, IRAS~04267+2600                                 			  & 04 29 51.6  &  26 06 44.8 	 &   140  \\
FX~Tau                  & HBC~44, Haro~6-11, IRAS~04274+2420                                       			  & 04 30 29.6  &  24 26 45.2 	 &   140  \\
DK~Tau                  & HBC~45, IRAS~04276+2554                                                  			  & 04 30 44.2  &  26 01 24.8 	 &   140  \\
V710~Tau                & HBC~51, LkH$\alpha$~266, IRAS~04290+1815                                 			  & 04 31 57.8  &  18 21 36.8 	 &   140  \\
GI~Tau                  & HBC~56, Haro~6-21, IRAS~04305+2414                                       			  & 04 33 34.1  &  24 21 17.0 	 &   140  \\
GK~Tau                  & HBC~57, Haro~6-22,                                                       			  & 04 33 34.6  &  24 21 05.9 	 &   140  \\
HN~Tau                  & HBC~60, Haro~6-24, St~31, IRAS~04307+1745                                                       & 04 33 39.4  &  17 51 52.3    &   140  \\
DM~Tau                  & HBC~62, IRAS~04309+1803                                                  			  & 04 33 48.7  &  18 10 10.0 	 &   140  \\
AA~Tau                  & HBC~63, IRAS~04318+2422                                                  			  & 04 34 55.4  &  24 28 53.2 	 &   140  \\
DN~Tau                  & HBC~65, IRAS~F04324+2408                                                 			  & 04 35 27.4  &  24 14 58.9 	 &   140  \\
CoKu Tau 3              & HBC 411                                                                  			  & 04 35 40.9  &  24 11 08.5    &   140  \\
DO~Tau                  & HBC~67, ITG~7                                                            			  & 04 38 28.6  &  26 10 49.9 	 &   140  \\
CoKu Tau 4              & HBC 421                                                                  			  & 04 41 16.8  &  28 40 00.5 	 &   140  \\
DP~Tau                  & HBC~70, IRAS~04395+2509                                                  			  & 04 42 37.7  &  25 15 37.5 	 &   140  \\
UY~Aur                  & HBC~76, IRAS~04486+3042                                                  			  & 04 51 47.4  &  30 47 13.9 	 &   140  \\
GM~Aur                  & HBC~77, IRAS~04519+3017                                                  			  & 04 55 11.0  &  30 21 59.4 	 &   140  \\
V836~Tau                & HBC~429, IRAS~C05000+2519                                                                       & 05 03 06.6  &  25 23 19.6    &   140  \\									      
RW~Aur                  & HBC~80, HD~240764, IRAS~04307+1745                                                              & 05 07 49.6  &  30 24 05.2    &   140  \\									      
BF~Ori                  & HBC~169, Haro~4-229, IRAS~05348-0636, Par~2510                           			  & 05 37 13.3  & -06 35 00.6 	 &   400  \\
IRAS 08267-3336         & HBC~562, AT~Pyx                                                          			  & 08 28 40.7  & -33 46 22.4 	 &   400  \\
WX~Cha                  & HBC~581, Sz~35, BYB~48, Ced~111~IRS~6, ISO-ChaI~228, IRAS~11085-7720     			  & 10 09 58.8  & -77 37 08.8 	 &   178  \\
SX~Cha                  & HBC~564, Sz~2, BYB~1                                                     			  & 10 55 59.7  & -77 24 40.0 	 &   178  \\
SY~Cha                  & HBC~565, Sz~3, CHX~1, BYB~2, IRAS~10552-7655                             			  & 10 56 30.5  & -77 11 39.5 	 &   178  \\
TW~Cha                  & HBC~567, BYB~5, CHXR~5, IRAS~10577-7706                                  			  & 10 59 01.1  & -77 22 40.8    &   178  \\
TW~Hya                  & HBC~568, TWA~1, IRAS~10594-3426                                          			  & 11 01 51.9  & -34 42 17.0 	 &    57  \\
CS~Cha                  & HBC~569, Sz~9, BYB~10, CHXR~10, IRAS~11011-7717                          			  & 11 02 25.1  & -77 33 36.0 	 &   178  \\
BYB~35                  & ISO-ChaI~101                                                             			  & 11 07 21.5  & -77 22 11.7 	 &   178  \\
CHXR~30                 & ISO-ChaI~122, ChaI~608                                                                          & 11 08 00.3  & -77 17 32.0    &   178  \\									      
VW~Cha                  & HBC~575, Sz~24, ISO-ChaI~123, CHXR~31                                  			  & 11 08 01.5  & -77 42 28.8 	 &   178  \\
VZ~Cha                  & HBC~578, Sz~31, CHXR~39, IRAS~11078-7607, Ced~112 IRS~1                  			  & 11 09 23.8  & -76 23 20.8 	 &   178  \\
ISO-ChaI~237            & ...                                                                      			  & 11 10 11.4  & -76 35 29.0 	 &   178  \\
Ced 111 IRS 7           & BYB~50, CHXR~47, IRAS~11091-7716                                         			  & 11 10 38.0  & -77 32 39.9 	 &   178  \\
HM~27                   & HBC~584, Sz~37, ISO-ChaI~254, IRAS 11093-7701(/10)                       			  & 11 10 49.6  & -77 17 51.7    &   178  \\
XX~Cha                  & HBC~586, Sz~39, CHXR~49                                                 			  & 11 11 39.7  & -76 20 15.3 	 &   178  \\
RX~J1111.7-7620         & CHX~18N                                                                  			  & 11 11 46.3  & -76 20 09.2 	 &   163  \\
T~Cha                   & HBC~591, IRAS~11547-7904, RX~J1157.2-7921                                			  & 11 57 13.5  & -79 21 31.3 	 &    66  \\
IRAS~12535-7623         & RX~J1257.1-7639, SSTc2d~J125711.7-764011                                 			  & 12 57 11.8  & -76 40 11.6 	 &   178  \\
Sz~50                   & ISO-ChaII~52, SSTc2d J130055.3-771022, RX J1300.9-7710                   			  & 13 00 55.4  & -77 10 22.1 	 &   178  \\
ISO-ChaII~54            & IRAS~F12571-7657, RX~J1301.0-7713                                        			  & 13 00 59.2  & -77 14 02.8 	 &   178  \\
DL~Cha                  & IRAS~13022-7650                                                          			  & 13 06 08.4  & -77 06 27.4 	 &   178  \\
HT~Lup                  & HBC~248, Sz~68, IRAS~15420-3408                                          			  & 15 45 12.9  & -34 17 30.5 	 &   145  \\
GW~Lup                  & HBC~249, Sz~71, IRAS~15435-3421                                          			  & 15 46 44.7  & -34 30 35.5 	 &   100  \\
Sz~73                   & HBC~600                                                                  			  & 15 47 57.0  & -35 14 35.2 	 &   100  \\
GQ~Lup                  & HBC~250, Sz~75, IRAS~15459-3529                                          			  & 15 49 12.1  & -35 39 04.0 	 &   100  \\
IM~Lup                  & HBC~605, Sz~82, IRAS~15528-3747                                          			  & 15 56 09.2  & -37 56 05.9 	 &   140  \\
RU~Lup                  & HBC~251, Sz~83, HD~142560                                                			  & 15 56 42.3  & -37 49 15.6 	 &   140  \\
RY~Lup                  & HBC~252, IRAS~15561-4013                                                 			  & 15 59 28.4  & -40 21 51.1 	 &   150  \\
EX~Lup                  & HBC~253, HD~325367, IRAS~15597-4010                                      			  & 16 03 05.5  & -40 18 24.8 	 &   150  \\
Sz~102                  & HBC~617, V1190~Sco, HH~228, IRAS~16051-3855, Krautter's Star             			  & 16 08 29.7  & -39 03 11.2 	 &   200  \\
\hline                  
\end{tabular}           
\label{tableA1}            
\normalsize             
\end{table*}

\setcounter{table}{0}
\begin{table*}
\caption{Targets (continued)}
\begin{tabular}{llrrr}
\hline
\hline
Star                    & Alternative                                                                                     &RA(J2000.0)& dec(J2000.0)     & dist.	\\					 
                        & names                                                                    			  & (h m s)  & (deg $\arcmin$ $\arcsec$)  & (pc)        \\
\hline
AS~205                  & HBC~254+632, V866~Sco, IRAS~16086-1830                                   			  & 16 11 31.3  & -18 38 26.2 	 &   120  \\
PZ99 J161411            & 1RXS~J161410.6-230542                                                    			  & 16 14 11.1  & -23 05 35.8 	 &   145  \\
Haro 1-1                & HBC~256, V895~Sco                                                        			  & 16 21 34.7  & -26 12 27.0 	 &   125  \\
Haro 1-4                & HBC~257, V2503~Oph, DoAr~16, IRAS~16221-2312                             			  & 16 25 10.5  & -23 19 14.5 	 &   125  \\
DoAr~24E                & HBC~639, ISO-Oph~36, ROX~C04                                             			  & 16 26 23.4  & -24 21 00.0    &   125  \\
DoAr~25                 & ISO-Oph~38, ROXN~1, ROX~C03, YLW~34, IRAS~16234-2436                                            & 16 26 23.7  & -24 43 14.0    &   125  \\									      
SR~21                   & ...                                                                      			  & 16 27 10.3  & -24 19 12.4 	 &   125  \\
IRS~51                  & ROXN~66, ISO-Oph~167, IRAS~16246-2436, YLW~45                            			  & 16 27 39.8  & -24 43 15.2 	 &   125  \\
SR~9                    & HBC~264, V2129~Oph, DoAr~34, ISO-Oph~168, ROX~29, IRAS 16246-2415        			  & 16 27 40.3  & -24 22 04.1 	 &   125  \\
V853~Oph                & HBC~266, DoAr~40, ROX~34, IRAS~16257-2421                                			  & 16 28 45.3  & -24 28 18.8 	 &   125  \\
ROX~42C                 & IRAS 16282-2427                                                          			  & 16 31 15.7  & -24 34 01.9 	 &   125  \\
ROXs~43A                & NTTS~162819-2423S                                                        			  & 16 31 20.1  & -24 30 05.5 	 &   125  \\
IRS~60                  & IRAS~16284-2418                                                          			  & 16 31 30.9  & -24 24 39.6 	 &   125  \\
Haro 1-16               & HBC~268, V2062~Oph, DoAr~44                                              			  & 16 31 33.5  & -24 27 37.1 	 &   125  \\
Haro 1-17               & HBC~648, DoAr~52                                                         			  & 16 32 21.9  & -24 42 14.9 	 &   125  \\
RNO~90                  & HBC~649, V2132~Oph, V1003~Oph, IRAS~16312-1542                           			  & 16 34 09.2  & -15 48 16.9 	 &   140  \\
Wa~Oph 6                & HBC~653, V2508~Oph, IRAS~16459-1411                                      			  & 16 48 45.6  & -14 16 35.8 	 &   140  \\
V1121~Oph               & HBC~270, AS~209, IRAS~16464-1416                                         			  & 16 49 15.3  & -14 22 08.8 	 &   125  \\
VV~Ser                  & HBC~282, IRAS~18262+0006                                                 			  & 18 28 47.9  &  00 08 39.8 	 &   260  \\
SSTc2d~J182900.8$^a$    & ...                                                                      			  & 18 29 00.9  &  00 29 31.6 	 &   260  \\
SSTc2d~J182909.8$^b$    & ...                                                                      			  & 18 29 09.8  &  00 34 45.8 	 &   260  \\
SSTc2d~J182928.2$^c$    & IRAS~18268-0025                                                          			  & 18 29 28.2  & -00 22 57.4 	 &   260  \\
EC~74                   & CK~9                                                                     			  & 18 29 55.7  &  01 14 31.6 	 &   260  \\
EC~82                   & CK~3                                                                     			  & 18 29 56.9  &  01 14 46.7 	 &   260  \\
EC~90                   & CK~1                                                                     			  & 18 29 57.7  &  01 14 06.0    &   260  \\
EC~92                   & ...                                                                      			  & 18 29 57.9  &  01 12 51.5 	 &   260  \\
CK~4                    & EC~97                                                                    			  & 18 29 58.2  &  01 15 21.6 	 &   260  \\
LkH$\alpha$~348         & HBC~285,IRAS~18316-0028                                                  			  & 18 34 12.6  & -00 26 21.8 	 &   260  \\
RX~J1842.9-3542         & ...                                                                      			  & 18 42 57.9  & -35 32 42.7 	 &   140  \\
RX~J1852.3-3700         & IRAS~18489-3703                                                          			  & 18 52 17.3  & -37 00 12.0 	 &   140  \\
\hline
\end{tabular}\\
$^a$ full name SSTc2d J182900.8+002931   \\
$^b$ full name SSTc2d J182909.8+003446   \\
$^c$ full name SSTc2d~J182928.2+002257 = IRAS 18268-0025 (note error in designation as published in \citealt{lahuis07}; dec(2000) is $<$ 0).   \\
\normalsize
\end{table*}

\begin{table*}
\caption{Mid-IR and X-ray observations}
\begin{tabular}{lrrrrr}
\hline
\hline
Star                    &{\it Spitzer} & X-ray       & X-ray start time$^c$  &  X-ray stop time$^c$  & Total X-ray    \\
                        & AOR$^a$      & ObsID$^b$   & y-m-d h:m:s           &  y-m-d h:m:s	     & exposure$^c$ (s) 	 	   \\
\hline
RNO~15                  & 5633280      & 0503670101  & 2007-07-31\ 04:55:25  & 2007-07-31\ 15:32:31  & 33135	     \\ 	      
LkH$\alpha$~270         & 5634048      & 0065820101  & 2002-02-27\ 22:48:25  & 2002-02-28\ 12:41:45  & 44808	    \\ 
LkH$\alpha$~271         & 11827968     & 0065820101  & 2002-02-27\ 22:48:25  & 2002-02-28\ 12:41:45  & 44808	     \\       
LkH$\alpha$~326         & 5634304      & ...	     & ...		     & ...		     & ...	    \\       
LkH$\alpha$~327         & 5634560      & ...	     & ...		     & ...		     & ...	    \\       
LkH$\alpha$~330         & 5634816      & ...	     & ...		     & ...		     & ...	    \\       
IRAS 03446+3254         & 5635072      & ...	     & ...		     & ...		     & ...	    \\         
BP~Tau                  & 14548224     & 0200370101  & 2004-08-15\ 06:36:51  & 2004-08-16\ 18:42:57  & 116334       \\
FM~Tau                  & 15119872     & 0203542001  & 2004-09-12\ 07:21:01  & 2004-09-12\ 15:47:38  & 26760        \\	
T~Tau N                 & (VLT)	       & 0301500101  & 2005-08-15\ 14:14:33  & 2005-08-16\ 12:55:22  & 65810	    \\
T~Tau S                 & (VLT)	       & 0301500101  & 2005-08-15\ 14:14:33  & 2005-08-16\ 12:55:22  & 65810	    \\
LkCa~8                  & 9832960      & ...	     & ...		     & ...		     & ...	      \\ 
DG~Tau                  & 14547968     & 0203540201  & 2004-08-17\ 06:30:29  & 2004-08-17\ 17:27:47  & 35302        \\
IQ~Tau                  & 9832704      & 0203541401  & 2005-02-09\ 03:23:22  & 2005-02-09\ 12:25:37  & 28946	    \\
FX~Tau                  & 9832448      & 0203541301  & 2004-08-25\ 11:12:21  & 2004-08-25\ 20:00:17  & (M2) 31251   \\  
DK~Tau                  & 14548736     & 0203541401  & 2005-02-09\ 03:23:22  & 2005-02-09\ 12:25:37  & 28946        \\
V710~Tau                & 5636608      & 0109060301  & 2000-09-09\ 19:10:41  & 2000-09-10\ 10:18:12  & 48922	    \\
GI~Tau                  & 14550528     & 0203540401  & 2005-02-21\ 01:40:40  & 2005-02-21\ 10:26:17  & 27549        \\
GK~Tau                  & 14550528     & 0203540401  & 2005-02-21\ 01:40:40  & 2005-02-21\ 10:26:17  & 27549        \\
HN~Tau                  & 3532032      & 0401870201  & 2007-02-12\ 17:53:16  & 2007-02-13\ 03:20:33  & 30070	    \\       
DM~Tau                  & 3536384      & 0554770101  & 2009-02-10\ 02:53:12  & 2009-02-10\ 13:43:49  & 34154        \\
AA~Tau                  & 14551552     & 0152680201  & 2003-02-14\ 02:40:58  & 2003-02-14\ 06:56:35  & 13719        \\
DN~Tau                  & 9831936      & 0203542101  & 2005-03-04\ 20:44:43  & 2005-03-05\ 04:56:58  & 26387	    \\
CoKu Tau 3              & 9831936      & 0203542101  & 2005-03-04\ 20:44:43  & 2005-03-05\ 04:56:58  & 26387	    \\
DO~Tau                  & 14548480     & 0501500101  & 2008-02-22\ 18:02:57  & 2008-02-23\ 01:54:48  & 24912        \\
CoKu Tau 4              & 5637888      & ...	     & ...		     & ...		     & ...	      \\	 
DP~Tau                  & (Gemini N)   & 0203542201  & 2005-03-05\ 06:18:52  & 2005-03-05\ 14:39:35  & 26919        \\
UY~Aur                  & 14551040     & 0401870501  & 2007-03-19\ 14:50:12  & 2007-03-20\ 01:04:21  & 31899        \\
GM~Aur                  & 18015488     & 0502100101  & 2007-09-07\ 08:53:53  & 2007-09-07\ 12:26:11  & 11048        \\
V836~Tau                & 3544320      & (CXO) 9300  & 2008-04-07\ 04:22:00  & 2008-04-07\ 06:58:58  & 8034	    \\
RW~Aur                  & 3534080      & 0401870301  & 2007-02-21\ 12:10:45  & 2007-02-21\ 21:46:22  & 30538	    \\       
BF~Ori                  & 5638144      & 0503560601  & 2007-09-08\ 08:44:17  & 2007-09-08\ 23:45:15  & 47648	     \\       
IRAS 08267-3336         & 5639168      & 0550120501  & 2008-06-08\ 06:26:06  & 2008-06-08\ 09:44:43  &  10507	    \\ 
WX~Cha                  & 5640192      & 0002740501  & 2002-04-09\ 10:20:23  & 2002-04-09\ 18:59:59  & 27941	     \\       
SX~Cha                  & 5639424      & 0152460301  & 2003-08-18\ 05:54:03  & 2003-08-18\ 14:25:13  & (M) 30371    \\       
SY~Cha                  & 5639424      & 0067140201  & 2001-02-24\ 05:50:17  & 2001-02-24\ 15:40:41  & 32065	    \\ 
TW~Cha                  & 5639680      & 0152460301  & 2003-08-18\ 06:15:25  & 2003-08-18\ 14:19:20  & 26301	   \\	    
TW~Hya                  & (Gemini N)   & 0112880201  & 2001-07-09\ 06:35:38  & 2001-07-09\ 13:57:31  & 23957        \\ 
CS~Cha                  & 12695808     & 0152460301  & 2003-08-18\ 06:15:25  & 2003-08-18\ 14:19:20  & 26301        \\ 
BYB~35                  & 5639680      & 0203810201  & 2004-02-27\ 16:48:11  &  2004-02-28\ 00:48:47 & (+M12) 26015 \\
CHXR~30                 & 18020096     & 0203810201  & 2004-02-27\ 16:48:11  & 2004-02-28\ 00:48:47  & 26015	    \\       
VW~Cha                  & 5639680      & 0002740501  & 2002-04-09\ 10:20:23  & 2002-04-09\ 18:59:59  & 27941	    \\ 
VZ~Cha                  & 5640448      & 0300270201  & 2005-09-02\ 03:33:56  & 2005-09-03\ 14:34:33  & 110399	    \\ 
ISO-ChaI~237            & 5640448      & 0203810101  & 2004-09-28\ 19:10:05  & 2004-09-29\ 02:24:04  & 23327	     \\       
Ced 111 IRS 7           & 5640192      & 0002740501  & 2002-04-09\ 10:20:23  & 2002-04-09\ 18:59:59  & 27941	    \\ 
HM~27                   & 5640192      & ...	     & ...		     & ...		     & ...	    \\       
XX~Cha                  & 5640448      & 0300270201  & 2005-09-02\ 03:33:56  & 2005-09-03\ 14:34:33  & 110399	    \\ 
RX~J1111.7-7620         & 5451776      & 0300270201  & 2005-09-02\ 03:33:56  & 2005-09-03\ 14:34:33  & 110399	    \\ 
T~Cha                   & 5641216      & 0550120601  & 2009-03-16\ 16:51:24  & 2009-03-16\ 18:31:58  & 5282	      \\       
IRAS~12535-7623         & 11827456     & ...	     & ...		     & ...		     & ...	    \\       
Sz~50                   & 11827456     & ...	     & ...		     & ...		     & ...	      \\       
ISO-ChaII~54            & 15735040     & ...	     & ...		     & ...		     & ...	      \\ 
DL~Cha                  & 5642240      & ...	     & ...		     & ...		     & ...	    \\       
HT~Lup                  & 5643264$^d$  & ...	     & ...		     & ...		     & ...	    \\       
GW~Lup                  & 5643520      & 0550120101  & 2008-08-21\ 09:04:36  & 2008-08-21\ 12:02:22  & (+M12) 9122  \\   
Sz~73                   & 5644032      & ...	     & ...		     & ...		     & ...	      \\       
GQ~Lup                  & 5644032      & 0550120301  & 2008-08-16\ 04:30:31  & 2008-08-16\ 06:40:27  & 6862	      \\	 
IM~Lup                  & 5644800      & 0303900401  & 2005-08-17\ 09:57:15  & 2005-08-17\ 16:52:52  & 21603	    \\ 
RU~Lup                  & 5644800      & 0303900401  & 2005-08-17\ 09:57:15  & 2005-08-17\ 16:52:52  & 21603	    \\ 
RY~Lup                  & 5644544      & ...	     & ...		     & ...		     & ...	    \\       
EX~Lup                  & 5645056      & 0503590101  & 2007-08-13\ 11:51:48  & 2007-08-13\ 16:02:18  & 13245	     \\       
Sz~102                  & 9407488      & 0146310201  & 2003-09-06\ 06:34:08  & 2003-09-06\ 12:23:53  & 18933	    \\ 
\hline                  
\end{tabular}           
\label{tableA2}            
\normalsize             
\end{table*}

\setcounter{table}{1}
\begin{table*}
\caption{Mid-IR and X-ray observations (continued)}
\begin{tabular}{lrrrrr}
\hline
\hline
Star                    & {\it Spitzer}& X-ray       & X-ray start time$^c$  &  X-ray stop time$^c$  & Total X-ray    \\
                        & AOR$^a$      & ObsID$^b$   & y-m-d h:m:s           &  y-m-d h:m:s	     & exposure$^c$ (s)  \\
\hline
AS~205                  & 5646080      & ...	     & ...		     & ...		     & ...	    \\       
PZ99 J161411            & 5453824      & 0109060201  & 2000-08-24\ 21:04:08  & 2000-08-25\ 11:47:38  & 47235	    \\ 
Haro 1-1                & 9833472      & ...	     & ...		     & ...		     & ...	    \\         
Haro 1-4                & 9833216      & ...	     & ...		     & ...		     & ...	    \\         
DoAr~24E                & 5647616      & (CXO) 637   & 2000-05-15\ 23:35:17  & 2000-05-17\ 03:18:30  & 96443	     \\
DoAr~25                 & 12663808     & 0305540501  & 2005-03-08\ 17:52:29  & 2005-03-09\ 23:06:18  & 83963	    \\       
SR~21                   & 5647616      & (CXO) 637   & 2000-05-15\ 23:35:17  & 2000-05-17\ 03:18:30  & 96443	     \\       
IRS~51                  & 9829888      & 0305540701  & 2005-03-12\ 14:36:55  & 2005-03-12\ 16:12:36  & (M) 56628    \\ 
SR~9                    & 12027392     & ...	     & ...		     & ...		     & ...	      \\ 
V853~Oph                & 12408576     & (CXO) 622   & 2000-07-04\ 10:13:29  & 2000-07-04\ 11:55:10  & 4812	      \\	 
ROX~42C                 & 6369792      & 0550120201  & 2008-08-20\ 22:39:41  & 2008-08-21\ 03:38:34  & 15773	    \\ 
ROXs~43A                & 15914496     & 0550120201  & 2008-08-20\ 22:39:41  & 2008-08-21\ 03:38:34  & 15773	     \\       
IRS~60                  & 6370048      & 0550120201  & 2008-08-20\ 22:39:41  & 2008-08-21\ 03:38:34  & 15773	    \\
Haro 1-16               & 12664064     & 0550120201  & 2008-08-20\ 22:39:41  & 2008-08-21\ 03:38:34  & 15773	    \\ 
Haro 1-17               & 11827712     & 0550120201  & 2008-08-20\ 22:39:41  & 2008-08-21\ 03:38:34  & 15773	    \\ 
RNO~90                  & 5650432      & ...	     & ...		     & ...		     & ...	    \\       
Wa~Oph 6                & 5650688      & ...	     & ...		     & ...		     & ...	    \\         
V1121~Oph               & 5650688      & ...	     & ...		     & ...		     & ...	    \\         
VV~Ser                  & 5651200      & ...	     & ...		     & ...		     & ...	    \\       
SSTc2d~J182900.8        & 13210112     & 0402820101  & 2007-04-15\ 13:52:40  & 2007-04-16\ 05:56:38  & 50802	 \\	  
SSTc2d~J182909.8        & 13210624     & 0402820101  & 2007-04-15\ 13:52:40  & 2007-04-16\ 05:56:38  & 50802	 \\	  
SSTc2d~J182928.2        & 13210368     & 0550120401  & 2008-09-19\ 00:58:20  & 2008-09-19\ 08:28:54  & 23757	\\ 
EC~74                   & 9407232      & 0065820201  & 2003-09-23\ 03:29:45  & 2003-09-23\ 08:29:16  & 16235	    \\ 
EC~82                   & 9407232      & (CXO) 4479  & 2004-06-19\ 21:42:42  & 2004/06/20\ 23:05:05  & 88446	    \\ 
EC~90                   & 9828352      & 0065820201  & 2003-09-23\ 03:29:45  & 2003-09-23\ 08:29:16  & 16235	     \\       
EC~92                   & 9407232      & (CXO) 4479  & 2004-06-19\ 21:42:42  & 2004/06/20\ 23:05:05  & 88446	    \\
CK~4                    & 9407232      & (CXO) 4479  & 2004-06-19\ 21:42:42  & 2004/06/20\ 23:05:05  & 88446	     \\
LkH$\alpha$~348         & 9831424      & ...	     & ...		     & ...		     & ...	    \\       
RX~J1842.9-3542         & 5451521      & ...	     & ...		     & ...		     & ...          \\       
RX~J1852.3-3700         & 5452033      & ...	     & ...		     & ...		     & ...          \\       
\hline
\end{tabular}\\
Notes:\\
$^a$ {\it Spitzer} AOR = Astronomical Observation Request; for further details on observing setup, see references in Table~\ref{table3}. \\
$^b$ {\it XMM-Newton} observation identification number if not otherwise noted; {\it Chandra} ID if ``(CXO)'' added. \\
$^c$ Exposure start and stop times for the {\it XMM-Newton} EPIC pn camera if not otherwise noted; if (M1), (M2), (M) added, then MOS1, MOS2, or both
     MOS were used for analysis, and exposure times refer to those detectors; (+M12) indicates that MOS detectors were
     used additionally to PN for spectral analysis. Exposure times in last column are livetimes for CCD\#1 of detector and are only 
     indicative as intervals with high background were selectively and additionally flagged.\\ 
$^d$ Also Spitzer AOR~9829120 for HT~Lup.
\normalsize
\end{table*}

\begin{table*}[t!]
\caption{Fluxes and luminosities}
\begin{tabular}{lrrrrrrrl}
\hline
\hline
Star                     & $f_{\rm [Ne\,II]}^a$		         & $L_{\rm [Ne\,II]}$	   &$f_{\rm X, 0.3-10, abs}$&$L_{\rm X, 0.3-10, abs}$&$f_{\rm X, 0.3-10, unabs}$&$L_{\rm X, 0.3-10, unabs}$& $N_{\rm H}^b$  & Refs.$^c$\\ 
                         &  (erg~cm$^{-2}$~s$^{-1}$)		 &      (erg~s$^{-1}$)	   &(erg~cm$^{-2}$~s$^{-1}$)& (erg~s$^{-1}$)         &(erg~cm$^{-2}$~s$^{-1}$)  & (erg~s$^{-1}$)           &   ($10^{22}$)  &  \\
\hline
RNO~15                   & $<2.6(0.85)\phantom{ }\times 10^{-14} $ & $<1.9\times 10^{29}$  &  $3.7\times 10^{-13}$  & $2.8\times 10^{30}$  &  $1.1\times 10^{-12}$ &  $8.0\times 10^{30}$  & 4.6       & 1,X	 \\  
LkH$\alpha$~270          & $ 1.3(0.35)\phantom{ }\times 10^{-14} $ & $ 9.7\times 10^{28}$  &  $3.2\times 10^{-13}$  & $2.4\times 10^{30}$  &  $1.5\times 10^{-12}$ &  $1.1\times 10^{31}$  & 0.78      & 1,X	     \\  
LkH$\alpha$~271          & $<3.6(1.2)\phantom{0}\times 10^{-15}  $ & $<2.7\times 10^{28}$  &  $3.8\times 10^{-14}$  & $2.8\times 10^{29}$  &  $1.1\times 10^{-13}$ &  $7.9\times 10^{29}$  & 3.8       & 1,X	     \\  
LkH$\alpha$~326          & $ 3.1(1.4)\phantom{0}\times 10^{-15}  $ & $ 2.3\times 10^{28}$  &  ...		    & ...                  &  ...	           &  ...                  & ...       & 1	     \\  
LkH$\alpha$~327          & $ 7.8(3.8)\phantom{0}\times 10^{-15}  $ & $ 5.8\times 10^{28}$  &  ...		    & ...                  &  ...	           &  ...                  & ...       & 1	     \\  
LkH$\alpha$~330          & $ 3.8(1.9)\phantom{0}\times 10^{-15}  $ & $ 2.8\times 10^{28}$  &  ...		    & ...                  &  ...	           &  ...                  & ...       & 1	     \\  
IRAS 03446+3254          & $ 4.3(0.99)\phantom{ }\times 10^{-15} $ & $ 3.2\times 10^{28}$  &  ...		    & ...                  &  ...	           &  ...                  &   ...     & 1,X	     \\  
BP~Tau                   & $ 2.9(0.4)\phantom{0}\times 10^{-15}  $ & $ 6.8\times 10^{27}$  &  $4.3\times 10^{-13}$  & $1.0\times 10^{30}$  &  $6.1\times 10^{-13}$ &  $1.4\times 10^{30}$  & 0.09      & 3,X	     \\  
FM~Tau                   & $<1.1(0.37)\phantom{ }\times 10^{-14} $ & $<2.6\times 10^{28}$  &  $1.5\times 10^{-13}$  & $3.5\times 10^{29}$  &  $2.2\times 10^{-13}$ &  $5.1\times 10^{29}$  & 0.15      & 3,X	     \\  
T~Tau N                  & $ 2.0(0.4)\phantom{0}\times 10^{-13}  $ & $ 4.7\times 10^{29}$  &  $1.9\times 10^{-12}$  & $4.4\times 10^{30}$  &  $4.0\times 10^{-12}$ &  $9.4\times 10^{30}$  & 0.34      & 5,X	     \\  
T~Tau S                  & $ 1.1(0.5)\phantom{0}\times 10^{-12}  $ & $ 2.6\times 10^{30}$  &  ...		    & ...                  &  $4.0\times 10^{-12}$ &  $9.4\times 10^{30}$  & 1.6$^d$   & 5,X	     \\  
LkCa~8                   & $ 5.0(0.93)\phantom{ }\times 10^{-15} $ & $ 1.2\times 10^{28}$  &  ...		    & ...                  &  ...	           &  ...                  & ...       & 1	     \\  
DG~Tau                   & $ 2.6(0.09)\phantom{ }\times 10^{-13} $ & $ 6.1\times 10^{29}$  &  $1.3\times 10^{-13}$  & $3.0\times 10^{29}$  &  $2.3\times 10^{-13}$ &  $5.5\times 10^{29}$  & 2.1$^e$   & 3,X	     \\  
IQ~Tau                   & $ 9.9(1.6)\phantom{0}\times 10^{-15}  $ & $ 2.3\times 10^{28}$  &  $8.3\times 10^{-14}$  & $2.0\times 10^{29}$  &  $1.4\times 10^{-13}$ &  $3.2\times 10^{29}$  & 0.35      & 1,X	     \\  
FX~Tau                   & $<5.7(1.9)\phantom{0}\times 10^{-15}  $ & $<1.3\times 10^{28}$  &  $8.2\times 10^{-14}$  & $1.9\times 10^{29}$  &  $1.7\times 10^{-13}$ &  $3.9\times 10^{29}$  & 0.20      & 1,X	     \\  
DK~Tau                   & $3.5(0.7)\phantom{0}\times 10^{-15}   $ & $ 8.2\times 10^{27}$  &  $2.0\times 10^{-13}$  & $4.6\times 10^{29}$  &  $3.6\times 10^{-13}$ &  $8.4\times 10^{29}$  & 0.24      & 3,X	     \\  
V710~Tau                 & $<2.4(0.8)\phantom{0}\times 10^{-15}  $ & $<5.6\times 10^{27}$  &  $2.5\times 10^{-13}$  & $5.9\times 10^{29}$  &  $5.0\times 10^{-13}$ &  $1.2\times 10^{30}$  & 0.23      & 1,X	     \\  
GI~Tau                   & $8.4(0.5)\phantom{0}\times 10^{-15}   $ & $ 2.0\times 10^{28}$  &  $1.6\times 10^{-13}$  & $3.8\times 10^{29}$  &  $2.8\times 10^{-13}$ &  $6.7\times 10^{29}$  & 0.31      & 3,X	     \\  
GK~Tau                   & $4.5(0.6)\phantom{0}\times 10^{-15}   $ & $ 1.1\times 10^{28}$  &  $2.3\times 10^{-13}$  & $5.4\times 10^{29}$  &  $5.3\times 10^{-13}$ &  $1.2\times 10^{30}$  & 0.40      & 3,X	     \\  
HN~Tau                   & $2.4(0.48)\phantom{ }\times 10^{-14}  $ & $ 5.6\times 10^{28}$  &  $8.0\times 10^{-14}$  & $1.9\times 10^{29}$  &  $1.4\times 10^{-13}$ &  $3.2\times 10^{29}$  & 1.1$^e$   & 3,X	     \\  
DM~Tau                   & $5.5(-)\phantom{00}\times 10^{-15}    $ & $ 1.3\times 10^{28}$  &  $5.7\times 10^{-13}$  & $1.3\times 10^{30}$  &  $8.4\times 10^{-13}$ &  $2.0\times 10^{30}$  & 0.14      & 6,X	     \\  
AA~Tau                   & $1.2(0.04)\phantom{ }\times 10^{-14}  $ & $ 2.8\times 10^{28}$  &  $2.0\times 10^{-13}$  & $4.6\times 10^{29}$  &  $4.4\times 10^{-13}$ &  $1.0\times 10^{30}$  & 0.93      & 3,X	     \\  
DN~Tau                   & $<4.2(1.4)\phantom{0}\times 10^{-15}  $ & $<9.9\times 10^{27}$  &  $3.7\times 10^{-13}$  & $8.7\times 10^{29}$  &  $4.6\times 10^{-13}$ &  $1.1\times 10^{30}$  & 0.058     & 1,X	     \\  
CoKu Tau 3               & $ 3.8(1.5)\phantom{0}\times 10^{-15}  $ & $ 8.9\times 10^{27}$  &  $1.1\times 10^{-12}$  & $2.6\times 10^{30}$  &  $2.4\times 10^{-12}$ &  $5.7\times 10^{30}$  & 0.38      & 1,X	     \\  
DO~Tau                   & $ 8.0(0.9)\phantom{0}\times 10^{-15}  $ & $ 1.9\times 10^{28}$  &  $3.1\times 10^{-14}$  & $7.3\times 10^{28}$  &  $1.0\times 10^{-13}$ &  $2.4\times 10^{29}$  & 0.63      & 3,X	     \\  
CoKu Tau 4               & $ 2.7(0.8)\phantom{0}\times 10^{-15}  $ & $ 6.3\times 10^{27}$  &  ...		    & ...                  &  ...	           &  ...                  &   ...     & 1	     \\  
DP~Tau                   & $<1.1(0.55)\phantom{ }\times 10^{-14} $ & $<2.6\times 10^{28}$  &  $1.8\times 10^{-14}$  & $4.2\times 10^{28}$  &  $4.2\times 10^{-14}$ &  $9.9\times 10^{28}$  & 3.8$^e$   & 4,X	     \\  
UY~Aur                   & $2.2(0.28)\phantom{ }\times 10^{-14}  $ & $ 5.2\times 10^{28}$  &  $1.1\times 10^{-13}$  & $2.6\times 10^{29}$  &  $1.7\times 10^{-13}$ &  $4.0\times 10^{29}$  & 0.12      & 3,X	     \\  
GM~Aur                   & $1.2(0.06)\phantom{ }\times 10^{-14}  $ & $ 2.8\times 10^{28}$  &  $5.1\times 10^{-13}$  & $1.2\times 10^{30}$  &  $6.8\times 10^{-13}$ &  $1.6\times 10^{30}$  & 0.11      & 3,X	     \\  
V836~Tau                 & $ 1.1(0.65)\phantom{ }\times 10^{-14} $ & $ 2.6\times 10^{28}$  &  $4.6\times 10^{-13}$  & $1.1\times 10^{30}$  &  $7.2\times 10^{-13}$ &  $1.7\times 10^{30}$  & 0.23      & 3,C	 \\  
RW~Aur                   & $<8.2(2.0)\phantom{0}\times 10^{-15}  $ & $<1.9\times 10^{28}$  &  $4.0\times 10^{-13}$  & $9.4\times 10^{29}$  &  $7.0\times 10^{-13}$ &  $1.6\times 10^{30}$  & 0.22      & 3,X	 \\  
BF~Ori                   & $<1.4(0.48)\phantom{ }\times 10^{-14} $ & $<2.7\times 10^{29}$  &  $1.4\times 10^{-14}$  & $2.7\times 10^{29}$  &  $1.7\times 10^{-13}$ &  $3.2\times 10^{30}$  & 0.04      & 1,X	     \\  
IRAS 08267-3336          & $ 8.3(1.2)\phantom{0}\times 10^{-15}  $ & $ 1.6\times 10^{29}$  &  $4.9\times 10^{-14}$  & $9.3\times 10^{29}$  &  $9.3\times 10^{-13}$ &  $1.8\times 10^{31}$  & 0.86      & 1,X	     \\  
WX~Cha                   & $<4.8(1.6)\phantom{0}\times 10^{-15}  $ & $<1.8\times 10^{28}$  &  $2.8\times 10^{-13}$  & $1.1\times 10^{30}$  &  $1.2\times 10^{-12}$ &  $4.6\times 10^{30}$  & 0.58      & 1,X	     \\  
SX~Cha                   & $<8.1(2.7)\phantom{0}\times 10^{-15}  $ & $<3.1\times 10^{28}$  &  $4.5\times 10^{-14}$  & $1.7\times 10^{29}$  &  $6.7\times 10^{-14}$ &  $2.6\times 10^{29}$  & 0.15      & 1,X	     \\  
SY~Cha                   & $ 2.6(1.0)\phantom{0}\times 10^{-15}  $ & $ 9.9\times 10^{27}$  &  $1.3\times 10^{-13}$  & $4.8\times 10^{29}$  &  $1.8\times 10^{-13}$ &  $6.9\times 10^{29}$  & 0.10      & 1,X	     \\  
TW~Cha                   & $<3.3(1.1)\phantom{0}\times 10^{-15}  $ & $<1.3\times 10^{28}$  &  $2.0\times 10^{-13}$  & $7.5\times 10^{29}$  &  $3.7\times 10^{-13}$ &  $1.4\times 10^{30}$  & 0.18      & 1,X	     \\  
TW~Hya                   & $5.9(1.1)\phantom{0}\times 10^{-14}   $ & $ 2.3\times 10^{28}$  &  $4.0\times 10^{-12}$  & $1.5\times 10^{30}$  &  $5.3\times 10^{-12}$ &  $2.1\times 10^{30}$  & 0.044     & 4,X	     \\  
CS~Cha                   & $4.3(-)\phantom{00}\times 10^{-14}    $ & $ 1.6\times 10^{29}$  &  $8.2\times 10^{-13}$  & $3.1\times 10^{30}$  &  $1.0\times 10^{-12}$ &  $3.8\times 10^{30}$  & 0.042     & 6,X	     \\  
BYB~35                   & $<1.2(0.4)\phantom{0}\times 10^{-15}  $ & $<4.6\times 10^{27}$  &  ...		    & ...                  & $<1.1\times 10^{-13}$ & $<4.3\times 10^{29}$  & 3.8$^f$   & 1,X	     \\  
CHXR~30                  & $<2.1(0.7)\phantom{0}\times 10^{-15}  $ & $<8.0\times 10^{27}$  &  $1.4\times 10^{-13}$  & $5.2\times 10^{29}$  &  $9.8\times 10^{-13}$ &  $3.7\times 10^{30}$  & 1.2       & 3,X	     \\  
VW~Cha                   & $ 3.7(0.33)\phantom{ }\times 10^{-14} $ & $ 1.4\times 10^{29}$  &  $6.4\times 10^{-13}$  & $2.4\times 10^{30}$  &  $2.2\times 10^{-12}$ &  $8.3\times 10^{30}$  & 0.48      & 1,X	     \\  
VZ~Cha                   & $ 3.8(1.70)\phantom{ }\times 10^{-15} $ & $ 1.4\times 10^{28}$  &  $6.6\times 10^{-14}$  & $2.5\times 10^{29}$  &  $1.4\times 10^{-13}$ &  $5.3\times 10^{29}$  & 0.21      & 1,X	     \\  
ISO-ChaI~237             & $<6.0(2.0)\phantom{0}\times 10^{-15}  $ & $<2.3\times 10^{28}$  &  $3.7\times 10^{-14}$  & $1.4\times 10^{29}$  &  $2.5\times 10^{-13}$ &  $9.4\times 10^{29}$  & 1.1       & 1,X	     \\  
Ced 111 IRS 7            & $<1.8(0.6)\phantom{0}\times 10^{-15}  $ & $<6.8\times 10^{27}$  &  $1.4\times 10^{-13}$  & $5.3\times 10^{29}$  &  $2.7\times 10^{-13}$ &  $1.0\times 10^{30}$  & 0.45      & 1,X	     \\  
HM~27                    & $<3.0(1.0)\phantom{0}\times 10^{-15}  $ & $<1.1\times 10^{28}$  &  ...		    & ...                  &  ...	           &  ...                  & ...       & 1	     \\  
XX~Cha                   & $ 5.0(1.1)\phantom{0}\times 10^{-15}  $ & $ 1.9\times 10^{28}$  &  $1.3\times 10^{-13}$  & $4.8\times 10^{29}$  &  $2.8\times 10^{-13}$ &  $1.1\times 10^{30}$  & 0.34      & 1,X	     \\  
RX~J1111.7-7620          & $ 5.1(1.2)\phantom{0}\times 10^{-15}  $ & $ 1.6\times 10^{28}$  &  $4.6\times 10^{-13}$  & $1.5\times 10^{30}$  &  $8.7\times 10^{-13}$ &  $2.8\times 10^{30}$  & 0.25      & 2,X	     \\  
T~Cha                    & $ 3.2(0.10)\phantom{ }\times 10^{-14} $ & $ 1.7\times 10^{28}$  &  $5.1\times 10^{-13}$  & $2.7\times 10^{29}$  &  $2.0\times 10^{-12}$ &  $1.1\times 10^{30}$  & 0.97      & 1,X	     \\  
IRAS~12535-7623          & $<3.9(1.3)\phantom{0}\times 10^{-15}  $ & $<1.5\times 10^{28}$  &  ...		    & ...                  &  ...	           &  ...                  & ...       & 1	     \\  
Sz~50                    & $ 2.8(0.77)\phantom{ }\times 10^{-16} $ & $ 1.1\times 10^{27}$  &  ...		    & ...                  &  ...	           &  ...                  & ...       & 1	     \\  
ISO-ChaII~54             & $ 5.6(0.91)\phantom{ }\times 10^{-15} $ & $ 2.1\times 10^{28}$  &  ...		    & ...                  &  ...	           &  ...                  & ...       & 1	     \\  
DL~Cha                   & $<6.3(2.1)\phantom{0}\times 10^{-13}  $ & $<2.4\times 10^{30}$  &  ...		    & ...                  &  ...	           &  ...                  & ...       & 1	     \\  
HT~Lup                   & $<2.9(0.96)\phantom{ }\times 10^{-14} $ & $<7.3\times 10^{28}$  &  ...		    & ...                  &  ...	           &  ...                  & ...       & 1	     \\  
GW~Lup                   & $<2.1(0.7)\phantom{0}\times 10^{-15}  $ & $<2.5\times 10^{27}$  &  $1.8\times 10^{-13}$  & $2.1\times 10^{29}$  &  $2.4\times 10^{-13}$ &  $2.9\times 10^{29}$  & 0.098     & 1,X	     \\  
Sz~73                    & $ 2.1(0.26)\phantom{ }\times 10^{-14} $ & $ 2.5\times 10^{28}$  &  ...		    & ...                  &  ...	           &  ...                  &...        & 1	     \\  
GQ~Lup                   & $<8.1(2.7)\phantom{0}\times 10^{-15}  $ & $<9.7\times 10^{27}$  &  $3.3\times 10^{-13}$  & $3.9\times 10^{29}$  &  $6.2\times 10^{-13}$ &  $7.4\times 10^{29}$  & 0.47      & 1,X	     \\  
IM~Lup                   & $ 9.8(1.5)\phantom{0}\times 10^{-15}  $ & $ 2.3\times 10^{28}$  &  $9.9\times 10^{-13}$  & $2.3\times 10^{30}$  &  $1.4\times 10^{-12}$ &  $3.2\times 10^{30}$  & 0.11      & 1,X	     \\  
RU~Lup                   & $ 2.6(0.83)\phantom{ }\times 10^{-14} $ & $ 6.1\times 10^{28}$  &  $2.7\times 10^{-13}$  & $6.4\times 10^{29}$  &  $4.3\times 10^{-13}$ &  $1.0\times 10^{30}$  & 0.14      & 1,X	     \\  
RY~Lup                   & $<9.6(3.2)\phantom{0}\times 10^{-15}  $ & $<2.6\times 10^{28}$  &  ...		    & ...                  &  ...	           &  ...                  & ...       & 1	     \\  
EX~Lup                   & $<6.6(2.2)\phantom{0}\times 10^{-15}  $ & $<1.8\times 10^{28}$  &  $4.7\times 10^{-13}$  & $1.3\times 10^{30}$  &  $5.5\times 10^{-13}$ &  $1.5\times 10^{30}$  & 0.033     & 1,X	       \\  
Sz~102                   & $ 3.6(0.15)\phantom{ }\times 10^{-14} $ & $ 1.7\times 10^{29}$  &  $8.6\times 10^{-15}$  & $4.1\times 10^{28}$  &  $3.8\times 10^{-14}$ &  $1.8\times 10^{29}$  & 0.24      & 1,X	     \\  
\noalign{\smallskip}\hline
\end{tabular}
\label{tableA3}            
\end{table*}

\setcounter{table}{2}
\begin{table*}[t!]
\caption{Fluxes and luminosities (continued)}
\begin{tabular}{lrrrrrrrl}
\hline
\hline
Star                     & $f_{\rm [Ne\,II]}^a$		         & $L_{\rm [Ne\,II]}$	   &$f_{\rm X, 0.3-10, abs}$&$L_{\rm X, 0.3-10, abs}$&$f_{\rm X, 0.3-10, unabs}$&$L_{\rm X, 0.3-10, unabs}$& $N_{\rm H}^b$  & Refs.$^c$\\ 
                         &  (erg~cm$^{-2}$~s$^{-1}$)	         &      (erg~s$^{-1}$)	   &(erg~cm$^{-2}$~s$^{-1}$)& (erg~s$^{-1}$)         &(erg~cm$^{-2}$~s$^{-1}$)  & (erg~s$^{-1}$)           &   ($10^{22}$)  &  \\
\hline
AS~205                   & $<8.1(2.7)\phantom{0}\times 10^{-14}  $ & $<1.4\times 10^{29}$  &  ...		    & ...                  &  ...	           &  ...                  & ...       & 1	     \\  
PZ99 J161411             & $ 6.2(1.8)\phantom{0}\times 10^{-15}  $ & $ 1.6\times 10^{28}$  &  $1.7\times 10^{-12}$  & $4.2\times 10^{30}$  &  $2.5\times 10^{-12}$ &  $6.2\times 10^{30}$  & 0.12      & 2,X	     \\  
Haro 1-1                 & $ 3.2(0.91)\phantom{ }\times 10^{-15} $ & $ 6.0\times 10^{27}$  &  ...		    & ...                  &  ...	           &  ...                  & ...       & 1	     \\  
Haro 1-4                 & $ 1.1(0.16)\phantom{ }\times 10^{-14} $ & $ 2.1\times 10^{28}$  &  ...		    & ...                  &  ...	           &  ...                  & ...       & 1	     \\  
DoAr~24E                 & $<3.6(1.2)\phantom{0}\times 10^{-14}  $ & $<6.7\times 10^{28}$  &  $1.0\times 10^{-13}$  & $1.9\times 10^{29}$  &  $3.0\times 10^{-13}$ &  $5.6\times 10^{29}$  & 1.2       & 1,C	     \\  
DoAr~25                  & $ 2.8(0.5)\phantom{0}\times 10^{-15}  $ & $ 5.2\times 10^{27}$  &  $2.3\times 10^{-13}$  & $4.3\times 10^{29}$  &  $1.5\times 10^{-12}$ &  $2.8\times 10^{30}$  & 1.5       & 3,X         \\  
SR~21                    & $<3.0(1.0)\phantom{0}\times 10^{-14}  $ & $<5.6\times 10^{28}$  &  $6.9\times 10^{-14}$  & $1.3\times 10^{29}$  &  $5.5\times 10^{-13}$ &  $1.0\times 10^{30}$  & 1.1       & 1,C	     \\  
IRS~51                   & $ 7.9(2.8)\phantom{0}\times 10^{-15}  $ & $ 1.5\times 10^{28}$  &  $2.3\times 10^{-13}$  & $4.3\times 10^{29}$  &  $1.8\times 10^{-12}$ &  $3.4\times 10^{30}$  & 3.7       & 1,X	     \\  
SR~9                     & $ 7.2(1.9)\phantom{0}\times 10^{-15}  $ & $ 1.3\times 10^{28}$  &  ...		    & ...                  &  ...	           &  ...                  & ...       & 1,X	     \\  
V853~Oph                 & $ 1.6(0.16)\phantom{ }\times 10^{-14} $ & $ 3.0\times 10^{28}$  &  $7.7\times 10^{-13}$  & $1.4\times 10^{30}$  &  $1.7\times 10^{-12}$ &  $3.1\times 10^{30}$  & 0.26      & 1,C	     \\
ROX~42C                  & $ 5.0(1.2)\phantom{0}\times 10^{-15}  $ & $ 9.4\times 10^{27}$  &  $1.1\times 10^{-12}$  & $2.1\times 10^{30}$  &  $2.4\times 10^{-12}$ &  $4.5\times 10^{30}$  & 0.39      & 1,X	     \\  
ROXs~43A                 & $<2.2(0.73)\phantom{ }\times 10^{-14} $ & $<4.1\times 10^{28}$  &  $2.2\times 10^{-12}$  & $4.2\times 10^{30}$  &  $5.0\times 10^{-12}$ &  $9.4\times 10^{30}$  & 0.39      & 1,X	     \\  
IRS~60                   & $ 1.4(0.42)\phantom{ }\times 10^{-14} $ & $ 2.6\times 10^{28}$  &  $8.5\times 10^{-14}$  & $1.6\times 10^{29}$  &  $1.4\times 10^{-13}$ &  $2.6\times 10^{29}$  & 0.41      & 1,X	     \\  
Haro 1-16                & $ 6.8(3.3)\phantom{0}\times 10^{-15}  $ & $ 1.3\times 10^{28}$  &  $2.9\times 10^{-13}$  & $5.3\times 10^{29}$  &  $6.8\times 10^{-13}$ &  $1.3\times 10^{30}$  & 0.35      & 1,X	     \\  
Haro 1-17                & $ 1.6(0.4)\phantom{0}\times 10^{-15}  $ & $ 3.0\times 10^{27}$  &  $1.6\times 10^{-13}$  & $3.0\times 10^{29}$  &  $3.1\times 10^{-13}$ &  $5.8\times 10^{29}$  & 0.25      & 1,X	     \\  
RNO~90                   & $ 2.4(0.84)\phantom{ }\times 10^{-14} $ & $ 5.6\times 10^{28}$  &  ...		    & ...                  &  ...	           &  ...                  & ...       & 1	     \\  
Wa~Oph 6                 & $ 9.1(3.8)\phantom{0}\times 10^{-15}  $ & $ 2.1\times 10^{28}$  &  ...		    & ...                  &  ...	           &  ...                  & ...       & 1	     \\  
V1121~Oph                & $ 1.9(0.69)\phantom{ }\times 10^{-14} $ & $ 3.6\times 10^{28}$  &  ...		    & ...                  &  ...	           &  ...                  & ...       & 1	     \\  
VV~Ser                   & $<3.9(1.3)\phantom{0}\times 10^{-14}  $ & $<3.2\times 10^{29}$  &  ...		    & ...                  &  ...	           &  ...                  & ...       & 1	     \\  
SSTc2d~J182900.8         & $<4.8(1.6)\phantom{0}\times 10^{-15}  $ & $<3.9\times 10^{28}$  &  $3.6\times 10^{-14}$  & $2.9\times 10^{29}$  &  $7.0\times 10^{-14}$ &  $5.7\times 10^{29}$  & 0.51      & 1,X	     \\  
SSTc2d~J182909.8         & $<1.3(0.44)\phantom{ }\times 10^{-14} $ & $<1.1\times 10^{29}$  &  (undet)		    & (undet)              &  (undet)		   &  (undet)              & ...       & 1,X	     \\  
SSTc2d~J182928.2         & $ 1.0(0.11)\phantom{ }\times 10^{-13} $ & $ 8.1\times 10^{29}$  &  (undet)		    & (undet)              &  (undet)		   &  (undet)              & ...       & 1,X	     \\  
EC~74                    & $ 4.2(1.1)\phantom{0}\times 10^{-15}  $ & $ 3.4\times 10^{28}$  &  $2.8\times 10^{-13}$  & $2.2\times 10^{30}$  &  $5.4\times 10^{-13}$ &  $4.3\times 10^{30}$  & 2.1       & 1,X	     \\  
EC~82                    & $ 7.1(2.6)\phantom{0}\times 10^{-15}  $ & $ 5.8\times 10^{28}$  &  $2.4\times 10^{-15}$  & $1.9\times 10^{28}$  &  $1.1\times 10^{-14}$ &  $9.2\times 10^{28}$  & 1.1       & 1,C	     \\  
EC~90                    & $<6.0(2.0)\phantom{0}\times 10^{-14}  $ & $<4.9\times 10^{29}$  &  $2.6\times 10^{-13}$  & $2.1\times 10^{30}$  &  $9.3\times 10^{-13}$ &  $7.6\times 10^{30}$  & 1.9       & 1,X	     \\  
EC~92                    & $ 1.5(0.12)\phantom{ }\times 10^{-14} $ & $ 1.2\times 10^{29}$  &  $1.7\times 10^{-13}$  & $1.4\times 10^{30}$  &  $1.2\times 10^{-12}$ &  $9.5\times 10^{30}$  & 5.6       & 1,C	     \\  
CK~4                     & $<3.6(1.2)\phantom{0}\times 10^{-15}  $ & $<2.9\times 10^{28}$  &  $1.2\times 10^{-14}$  & $9.6\times 10^{28}$  &  $8.8\times 10^{-14}$ &  $7.2\times 10^{29}$  & 1.4       & 1,C	     \\  
LkH$\alpha$~348          & $<6.6(2.2)\phantom{0}\times 10^{-14}  $ & $<5.4\times 10^{29}$  &  ...		    & ...                  &  ...	           &  ...                  & ...       & 1	     \\  
RX~J1842.9-3542          & $ 4.3(1.3)\phantom{0}\times 10^{-15}  $ & $ 1.0\times 10^{28}$  &  ...		    & ...                  &  ...	           &  ...                  & ...       & 2	     \\  
RX~J1852.3-3700          & $ 7.2(0.4)\phantom{0}\times 10^{-15}  $ & $ 1.7\times 10^{28}$  &  ...		    & ...                  &  ...	           &  ...                  & ...       & 2	     \\  
\noalign{\smallskip}\hline
\end{tabular}\\
Notes:\\
$^a$ Errors are given in parentheses.   \\
$^b$ In units of $10^{22}$~cm$^{-2}$.   \\
$^c$ References: 1: New analysis of Spitzer c2d sample, for original analysis see \citet{lahuis07}; 2: \citet{pascucci07}; 3: archival GTO/GO observations, see Najita \& Carr, in prep.;
                 4: \citet{herczeg07}; 5: \citet{boekel09}; 6: \citet{espaillat07}; C: from {\it Chandra} archive; X: from {\it XMM-Newton} archive.   \\
$^d$ Same intrinsic (unabsorbed) $L_{\rm X}$ as for T Tau N assumed (similar stellar masses). Minimum $N_{\rm H}$ based on minimum estimated visual 
     extinction of 8~mag \citep{duchene02}.  \\
$^e$ $N_{\rm H}$ and $f_{\rm X}$ only for hard (coronal) component in Two Absorber  X-ray spectra \citep{guedel07b, guedel09b}.  \\
$^f$ $N_{\rm H}$ adopted based on published $A_{\rm V}$ or $A_{\rm J}$ (see text for details).   \\
\end{table*}

\begin{table*}[t!]
\caption{Additional parameters}
\begin{tabular}{lrrrrrrrrr}
\hline
\hline
Star                     & $\log \dot{M}_{\rm acc}$ &  EW(H$\alpha$)  & EW([O\,{\sc i}]$_f$) &   EW([O\,{\sc i}]$_t$) & $\log L_{{\rm [OI]},_f}$  & $\log L_{{\rm [OI]},t}$  & $\log \dot{M}_{\rm loss}$    & jet? & Refs.$^b$   \\ 
                         & ($M_{\odot}$~yr$^{-1}$)  &      (\AA)      & (\AA)                &   (\AA)                & ($L_{\odot}$)             & ($L_{\odot}$)          & ($M_{\odot}$~yr$^{-1}$)   & TD?$^a$  &   \\
\hline
RNO~15                   &  ...   & ...    &     ...   &  ...   &   ...  &    ...  &    ...  &   ...  &  ...\\  
LkH$\alpha$~270          &  ...   &   30.9 &     ...   &  ...   &   ...  &    ...  &    ...  &   ...  &  6 \\  
LkH$\alpha$~271          &  ...   & 185.7  &      ...  &  ...   &   ...  &    ...  &    ...  &   ...  &  6\\  
LkH$\alpha$~326          &  ...   &   52.7 &     ...   & 0.8    &   ...  &    ...  &    ...  &   ...  &  6 \\  
LkH$\alpha$~327          &  ...   &   51.0 &     ...   & 0.9    &   ...  &    ...  &    ...  &   ...  &  6 \\  
LkH$\alpha$~330          &  ...   &   20.3 &     ...   &  ...   &   ...  &    ...  &    ...  &   T    &  6,31 \\  
IRAS 03446+3254          &  ...   &  ...   &     ...   & ...    &   ...  &    ...  &    ...  &   ...  &  ... \\  
BP~Tau                   & -7.54  &  40-92 &    0.07   &   0.26 &  -5.71 & -4.75   & $<$-9.7 &   ...  &  13,X,12 \\  
FM~Tau                   & -8.45  &  62-71 &     0.045 & 0.48   &  -6.37 &  -5.19  & $<$-10.6&   ...  &  21,35,6,12       \\  
T~Tau N                  & -7.12  &  38-60 &     ...   &  2.0   &   ...  &  -2.78  &  -6.96  &   J    &  21,6,26,4,(27) \\  
T~Tau S                  & ...    &   ...  &     ...   &  ...   &   ...  &    ...  &    ...  &   J    &  4 \\  
LkCa~8                   & -9.10  &  ...   &   0.24    & 0.63   & -5.3   &  -4.77  &   -9.6  &   ...  &  13,12 \\  
DG~Tau                   & -6.30  &  63-125&    12.4   &   17.5 &  -2.21 & -2.07   &   -6.5  &   J    &  21,X,12,29 \\  
IQ~Tau                   & -7.55  &  8-17  &     ...   &  0.4   &   ...  &  -4.89  &    ...  &   ...  &  21,6,(5,27) \\  
FX~Tau                   & -8.65  &  9.6   &      ...  &  ...   &   ...  &    ...  &    ...  &   ...  &  21,6\\  
DK~Tau                   & -7.42  &  19-50 &    0.36   &   0.79 &  -4.23 & -3.90   &   -8.5  &   ...  &  13,X,12 \\  
V710~Tau                 & ...    &  34-89 &    ...    & 0.3    &   ...  &  -5.00  &    ...  &   ...  &  X,6,(5,27) \\  
GI~Tau                   & -8.02  &  15-21 &    0.11   &   0.48 &  -5.11 & -4.46   &   -9.4  &   ...  &  13,X,12 \\  
GK~Tau                   & -8.19  &  15-35 &    0.20   &   0.51 &  -4.88 & -4.48   &   -9.2  &   ...  &  13,X,12 \\  
HN~Tau                   & -8.60  &120-163 &     5.7   & 13.2   &  -3.82 &  -3.46  &  -8.1   &   J    &  35,6,12,37   \\  
DM~Tau                   & -7.95  & 139    &    ...    &  ...   &   ...  &   ...   &    ...  &   T    &  13,6,33 \\  
AA~Tau                   & -8.48  &  37-46 &    0.16   &   1.25 &  -4.77 & -3.88   &   -9.1  &   J    &  13,X,12,30 \\  
DN~Tau                   & -8.46  &  12-87 &   0.1     & 0.53   &  -5.07 &  -4.34  &   -9.4  &   ...  &  21,X,12 \\  
CoKu Tau 3               &  ...   &    5   &     ...   &  ...   &   ...  &    ...  &    ...  &   ...  &  6 \\  
DO~Tau                   & -6.84  & 100-109&    2.63   &   3.60 &  -3.20 & -3.06   &   -7.5  &   J    &  13,X,12,17 \\  
CoKu Tau 4               &$<$-10.0&   ...  &     ...   &  ...   &   ...  &    ...  &    ...  &   T    &  21,34 \\  
DP~Tau                   & -7.88  & 74-102 &   0.7     &  0.7   &  -4.78 & -4.78   &  -7.42  &   J    &  13,X,17,26,28,(27) \\  
UY~Aur                   & -7.18  &  36-73 &    1.15   &   2.30 &  -3.91 & -3.62   &   -8.2  &   J    &  13,X,12,17 \\  
GM~Aur                   & -8.02  &  96-97 &   0.015   &   0.27 &  -6.25 & -4.90   &$<$-10.0 &   T    &  13,X,12,33 \\  
V836~Tau                 & -7.86  &  25    &     0.02  & 0.35   &  -6.60 &  -4.94  &$<$-10.1 &   T    &  26,12,38  \\  
RW~Aur                   & -7.12  &  42-84 &     1.18  & 1.59   &  -3.26 &  -3.11  &  -7.6   &   J    &  21,6,26,35,12,36     \\  
BF~Ori                   &  ...   &  6.4   &      ...  &  ...   &   ...  &    ...  &    ...  &   ...  &  6\\  
IRAS 08267-3336          &  ...   &  ...   &     ...   & ...    &   ...  &    ...  &    ...  &   ...  &  ... \\  
WX~Cha                   & -8.47  &  66    &      ...  &  ...   &   ...  &   ...   &    ...  &    ..  &  11,9\\  
SX~Cha                   & -8.37  &  27    &      ...  &  ...   &   ...  &   ...   &    ...  &   ...  &  13,9 \\  
SY~Cha                   & -8.60  &  8-64  &      ...  &  ...   &   ...  &    ...  &    ...  &   ...  &  13,20\&9 \\  
TW~Cha                   &  ...   & 26-41  &      ...  & 0.95   &   ...  &  -3.84  &    ...  &   ...  &  20\&9,10,(9)\\  
TW~Hya                   & -8.70  & 220    &    ...    & 0.67   &   ...  &  -5.06  &    ...  &   T    &  14,40,16,32 \\  
CS~Cha                   & -7.92  &  13-65 &    ...    &  ...   &   ...  &   ...   &    ...  &   J,T  &  8,9\&11,25,8 \\  
BYB~35                   &  ...   &    ... &  ...      &  ...   &   ...  &     ... &   ...   &   ...  &  ...\\  
CHXR~30                  &  ...   &  1     &      ...  & ...    &   ...  &   ...   &    ...  &  ...   &  20  \\  
VW~Cha                   & -6.95  &  71-147&      ...  & 0.93   &   ...  &  -3.63  &    ...  &   J    &  13,20\&9,10,3,(9) \\  
VZ~Cha                   & -8.28  &    71.4&      ...  & 0.42   &   ...  &  -4.75  &    ...  &   ...  &  13,9,10,(9) \\  
ISO-ChaI~237             &  ...   & 3.8    &      ...  &  ...   &   ...  &   ...   &    ...  &    ..  &  20 \\  
Ced 111 IRS 7            & ...    & ...    &    ...    &  ...   &   ...  &    ...  &    ...  &   ...  &  ... \\  
HM~27                    & ...    & 102-201&       ... &    ... &     ...&     ... &   ...   &     ...&   20\&9\\  
XX~Cha                   & -9.07  & 90-134 &      ...  &  ...   &   ...  &    ...  &    ...  &   ...  &  13,20\&9 \\  
RX~J1111.7-7620          & -9.30  &   7.7  &    ...    &  ...   &   ...  &    ...  &    ...  &    T   &  22,20 \\  
T~Cha                    &  ...   &  2-7   &      ...  &  ...   &   ...  &    ...  &    ...  &    T   &  41,31 \\  
IRAS~12535-7623          & ...    & ...    &       ... &    ... &     ...&     ... &    ...  &     ...&  ...\\  
Sz~50                    & ...    &   ...  &      ...  &  ...   &   ...  &    ...  &    ...  &    J   &  1 \\  
ISO-ChaII~54             &  ...   &  ...   &     ...   &  ...   &   ...  &    ...  &    ...  &   ...  &  ... \\  
DL~Cha                   & ...    & ...    &       ... &    ... &     ...&     ... &    ...  &   ...  &   ...\\  
HT~Lup                   & ...    &  2.8   &       ... & $<$0.03&  ...   & $<$-5.02&   ...   &   ...  &  18,10,(M,19)\\  
GW~Lup                   & ...    &  90.3  &    ...    &  ...   &   ...  &    ...  &    ...  &   ...  &  18 \\  
Sz~73                    &  ...   &   97.2 &      ...  &  ...   &   ...  &    ...  &    ...  &   ...  &  18 \\  
GQ~Lup                   & -7.50  &  38.6  &    ...    & 0.08   &   ...  &  -5.34  &    ...  &   ...  &  24,18,10,(M,19) \\  
IM~Lup                   & ...    &    8.1 &     ...   & ...    &   ...  &    ...  &    ...  &   ...  &  18 \\  
RU~Lup                   & -7.30  &  216.4 &     0.79  &   1.94 &   -4.3 &  -3.95  &    ...  &   J    &  15,18,10,25,(S,M,15) \\  
RY~Lup                   & ...    & ...    &       ... &    ... &     ...&     ... &    ...  &   ...  &   ...\\  
EX~Lup                   & ...    & ...    &       ... &    ... &     ...&     ... &    ...  &   ...  &   ...\\  
Sz~102                   & -8.10  &  377.4 &      ...  &  ...   &   ...  &    ...  &    ...  &   J    &  7,18,2 \\  
\noalign{\smallskip}\hline
\end{tabular}
\label{tableA4}            
\end{table*}

\setcounter{table}{3}
\begin{table*}[t!]
\caption{Additional parameters (continued)}
\begin{tabular}{lrrrrrrrrr}
\hline
\hline
Star                     & $\log \dot{M}_{\rm acc}$ &  EW(H$\alpha$)  & EW([O\,{\sc i}]$_f$) &   EW([O\,{\sc i}]$_t$) & $\log L_{{\rm [OI]},_f}$  & $\log L_{{\rm [OI]},t}$  & $\log \dot{M}_{\rm loss}$    & jet? & Refs.$^b$   \\ 
                         & ($M_{\odot}$~yr$^{-1}$)  &      (\AA)      & (\AA)                &   (\AA)                & ($L_{\odot}$)             & ($L_{\odot}$)            & ($M_{\odot}$~yr$^{-1}$)   & TD?$^a$  &   \\
\hline
AS~205                   & ...    & 154.6  &       ... &   1.8  &     ...&   3.16  &      ...&   ...  &   6,(M,6) \\  
PZ99 J161411             & -9.50  & ...    &    ...    &  ...   &   ...  &    ...  &    ...  &   T    &  22 \\  
Haro 1-1                 &  ...   &  123   &     ...   & 0.52   &   ...  &  -4.62  &    ...  &   ...  &  6,10,(S,6) \\  
Haro 1-4                 &  ...   &   47.6 &     ...   & ...    &   ...  &    ...  &    ...  &   ...  &  6 \\  
DoAr~24E                 & -8.21  & ...    &       ... &    ... &     ...&     ... &      ...&   ...  &  39 \\  
DoAr~25                  &$<$-9.24&  ...   &      ...  & ...    &   ...  &   ...   &    ...  &  ...   &  39 \\  
SR~21                    &$<$-8.84& ...    &       ... &    ... &     ...&     ... &      ...&   T    &  39,31 \\  
IRS~51                   &  ...   &  ...   &     ...   &  ...   &   ...  &    ...  &    ...  &   ...  &  ... \\  
SR~9                     & -8.26  &   11.6 &      ...  & 0.06   &   ...  &  -6.00  &    ...  &   ...  &  39,6,10,(S,6) \\  
V853~Oph                 & -8.31  &    42  &      ...  &  ...   &   ...  &    ...  &    ...  &   ...  &  39,23 \\
ROX~42C                  &  ...   &  ...   &     ...   &  ...   &   ...  &    ...  &    ...  &   ...  &  ... \\  
ROXs~43A                 & ...    & ...    &       ... &    ... &     ...&     ... &      ...&   ...  &  ...\\  
IRS~60                   &  ...   &  ...   &     ...   &  ...   &   ...  &    ...  &    ...  &   ...  &  ... \\  
Haro 1-16                &  ...   &   59   &     ...   & ...    &   ...  &    ...  &    ...  &   ...  &  6 \\  
Haro 1-17                &  ...   &   15   &    ...    &  ...   &   ...  &    ...  &    ...  &   ...  &  23 \\  
RNO~90                   &  ...   &  ...   &     ...   &  ...   &   ...  &    ...  &    ...  &   ...  &  ... \\  
Wa~Oph 6                 &  ...   &   ...  &      ...  &  ...   &   ...  &    ...  &    ...  &   ...  &  ... \\  
V1121~Oph                &  ...   &   ...  &      ...  &  ...   &   ...  &    ...  &    ...  &   ...  &  ... \\  
VV~Ser                   &  ...   &   ...  &         ..&  ...   &     ...&      ...&      ...&   ...  &  ...\\  
SSTc2d~J182900.8         &  ...   &    ... &    ...    &    ... &     ...&     ... &	 ... &   ...  &  ...\\  
SSTc2d~J182909.8         &  ...   &    ... &    ...    &    ... &     ...&     ... &	 ... &   ...  &  ...\\  
SSTc2d~J182928.2         &  ...   &  ...   &      ...  &  ...   &   ...  &    ...  &    ...  &   ...  &  ... \\  
EC~74                    &  ...   &  ...   &     ...   &  ...   &   ...  &    ...  &    ...  &   ...  &  ... \\  
EC~82                    &  ...   &  ...   &     ...   &  ...   &   ...  &    ...  &    ...  &   ...  &  ... \\  
EC~90                    &  ...   &    ... &    ...    &    ... &     ...&     ... &   ...   &   ...  &  ...\\  
EC~92                    &  ...   &  ...   &     ...   & ...    &   ...  &    ...  &    ...  &   ...  &  ... \\  
CK~4                     &  ...   &    ... &    ...    &    ... &     ...&       ..&   ...   &	 ...  &  ...\\  
LkH$\alpha$~348          &  ...   &    310 &    ...    &    0.6 &     ...&       ..&   ...   &	 ...  &  6  \\  
RX~J1842.9-3542          & -9.0   &    ... &    ...    &    ... &     ...&       ..&   ...   &	 T    &  22\\  
RX~J1852.3-3700          & -9.3   &    ... &    ...    &    ... &     ...&       ..&   ...   &	 T    &  22\\  
\noalign{\smallskip}\hline
\end{tabular}\\
Notes:\\
$^a$ T = transitional disk (either ``anemic'' or ``cold'' disk); J = jet-driving object.\\
$^b$ References:
    1 \citet{alcala08}; 
    2 \citet{bacciotti99};
    3 \citet{bally06};
    4 \citet{boehm94};
    5 \citet{briceno02};
    6 \citet{cohen79};
    7 \citet{comeron03}; 
    8 \citet{espaillat07};
    9 \citet{gauvin92};
    10 \citet{hamann94};
    11 \citet{hartigan93};
    12 \citet{hartigan95};
    13 \citet{hartmann98};
    14 \citet{herczeg04};
    15 \citet{herczeg05};
    16 \citet{herczeg07};
    17 \citet{hirth97};
    18 \citet{hughes94};
    19 \citet{krautter97};
    20 \citet{luhman04};
    21 \citet{najita07};
    22 \citet{pascucci07};
    23 \citet{rydgren80};
    24 \citet{seperuelo08};
    25 \citet{takami03};
    26 \citet{white04};
    27 \citet{kenyon95};
    28 \citet{mundt98};
    29 \citet{solf93};
    30 \citet{cox05};
    31 \citet{brown07};
    32 \citet{calvet02};
    33 \citet{calvet05};
    34 \citet{forrest04};
    35 \citet{white01};
    36 \citet{dougados00};
    37 \citet{hartigan04};
    38 \citet{najita08};
    39 \citet{natta06};
    40 \citet{webb99};
    41 \citet{alcala93}.
    M: 2MASS; 
    S: SIMBAD; 
    X: XEST survey, see \citet{guedel07a} and references therein.
    References in parentheses are for complementary data, e.g., $A_{\rm V}$ or R magnitude, used to calculate $\log L_{\rm [OI]}$.  \\
\end{table*}

\end{document}